\documentclass[journal,a4paper]{IEEEtran}   
\ifCLASSINFOpdf
\else
\usepackage[dvips]{graphicx,color}
\fi

\usepackage[cmex10]{amsmath}
\usepackage[tight,footnotesize]{subfigure}

\usepackage{amssymb}
\usepackage{epsfig}   
\usepackage{eufrak}

\newcommand{\mathcalP}{\mathfrak{P}}
\newcommand{\mathcalM}{\mathfrak{M}}
\begin{document}
%
\title{On Bregman Distances and Divergences of Probability Measures}
%
%
%

\author{Wolfgang~Stummer 
        and~Igor~Vajda,~\IEEEmembership{Fellow,~IEEE}
\thanks{W. Stummer is with the Department of Mathematics, 
University of Erlangen--N\"{u}rnberg, Cauerstrasse $11$, 
91058 Erlangen, Germany (e-mail: stummer@mi.uni-erlangen.de),
\ as well as with the School of Business and Economics,
University of Erlangen--N\"{u}rnberg.
}
\thanks{
I. Vajda is with the Institute of Information Theory and Automation, 
Academy of Sciences of the Czech Republic, 
Pod Vod\'{a}renskou V\v{e}\v{z}\'{\i} 4, 182\thinspace 08 Prague, Czech
Republic (e-mail: vajda@utia.cas.cz).
}
\thanks{Manuscript received October 26, 2009; revised August 4, 2011.
This work was supported by the M\v{S}MT grant 1M0572 and the GA\thinspace \v{C}R
grant 102/07/1131.
}\\
\thanks{\copyright 2011 IEEE. Personal use of this material is permitted. Permission from IEEE must be obtained for all other uses, in any current or future media, including reprinting/republishing this material for advertising or promotional purposes, creating new collective works, for resale or redistribution to servers or lists, or reuse of any copyrighted component of this work in other works.
}}

%
%

\markboth{
TO APPEAR IN IEEE TRANSACTIONS ON INFORMATION THEORY
}%
{ }
%



\maketitle

\begin{abstract}
The paper introduces scaled Bregman
distances of probability distributions which admit non-uniform contributions
of observed events. They are introduced in a general form covering not only
the distances of discrete and continuous stochastic observations, but  also
the distances of random processes and signals. It is shown that the scaled
Bregman distances extend not only the classical ones studied in the previous
literature, but also the information divergence and the related wider class
of convex divergences
of probability measures. An information processing theorem is
established too, but only in the sense of invariance w.r.t. statistically
sufficient transformations and not in the sense of universal monotonicity.
Pathological situations where coding can increase the classical Bregman
distance are illustrated by a concrete example. In addition to the classical
areas of application of the Bregman distances and convex divergences such as
recognition, classification, learning and evaluation of proximity of various
features and signals, the paper mentions a new application in 3D-exploratory
data analysis. Explicit expressions for the scaled Bregman distances are
obtained in general exponential families, with concrete applications in the
binomial, Poisson and Rayleigh families, and in the families of exponential
processes such as the Poisson and diffusion processes including the classical
examples of the Wiener process and geometric Brownian motion.
\end{abstract}

\vspace{0.1cm}

%
\IEEEpeerreviewmaketitle


\section{Introduction}

%
%
%
%


 

\IEEEPARstart{B}{regman} (1967) introduced
for convex functions $\phi :\mathbb{R}^{d}\rightarrow \mathbb{R}$ 
with gradient 
$\triangledown \phi $ the $\phi $-depending nonnegative measure of
dissimilarity
\begin{equation}
B_{\phi }(p,q)=\phi (p)-\phi (q)-\triangledown \phi (q)(p-q)  \label{B}
\end{equation}
of $d$-dimensional vectors $p,q\in \mathbb{R}^{d}$. His motivation was the
problem of convex programming, but in the subsequent literature it became
widely applied in many other problems under the name 
\textit{Bregman distance} in spite of that it is not in general the usual metric distance
(it is a pseudodistance which is reflexive but neither symmetric nor satisfying the
triangle inequality). The most important   feature is the special \textit{separable
form} defined by
\begin{equation}
B_{\phi }(p,q)=\sum_{i=1}^{d}\left[ \phi (p_{i})-\phi (q_{i})-\phi ^{\prime
}(q_{i})(p_{i}-q_{i})\right]  \label{b}
\end{equation}
for vectors $p=(p_{1},...,p_{d}),q=(q_{1},...,q_{d})$\ and convex
differentiable functions $\phi :\mathbb{R}\rightarrow \mathbb{R}$. For
example, the function $\phi (t)=(t-1)^{2}$ leads to the classical squared
Euclidean distance
\begin{equation}
B_{\phi }(p,q)=\sum_{i=1}^{d}\left( p_{i}-q_{i}\right) ^{2}.  \label{e}
\end{equation}
\medskip

In the optimization-theoretic context the Bregman distances are usually
studied in the general form (\ref{B}) -- see, e.g.,\ Csisz\'{a}r  and Mat\'{u}\v{s}
(2008, 2009), as well as Bauschke and Borwein (1997) for adjecent random projection studies. 
In the information-theoretic or statistical context they are
typically used in the separable form (\ref{b}) for vectors $p,q$\ with
nonnegative coordinates representing generalized distributions (finite
  \textit{discrete} measures) and functions $\phi :[0,\infty )\rightarrow \mathbb{R}$
differentiable on $(0,\infty )$ (the problem with $q_{i}=0$\ is solved by
resorting to the right-hand derivative $\phi _{+}^{\prime }(0)$). The
concrete example $\phi (t)=t\ln t$\ leads to the well-known Kullback
divergence 
\begin{equation}
B_{\phi }(p,q)=\sum_{i=1}^{d}p_{i}\ln \frac{p_{i}}{q_{i}}.  
\notag
\end{equation}
Of course, the 
most common context are discrete probability distributions $p,q$\ since
vectors of hypothetical or observed frequencies $p,q$\ are easily
transformed to the relative frequencies normed to $1$. For example, Csisz\'{a}r (1991, 1994, 1995) or Pardo  and Vajda (1997, 2003) used the Bregman
distances of probability distributions in the context of information theory
and asymptotic statistics.\medskip

Important alternatives to the Bregman distances (\ref{b}) are the $\phi $-
\textit{divergences} 
defined by
\begin{equation}
D_{\phi }(p,q)=\sum_{i=1}^{d}q_{i}\phi \left( \frac{p_{i}}{q_{i}}\right)
\label{5}
\end{equation}
for functions $\phi$ which are convex on $[0,\infty )$, 
continuous on $(0,\infty )$ and strictly convex at $1$\ with $\phi (1)=0$.
Originating in the paper of Csisz\'{a}r (1963), they share some properties
with the Bregman distances (\ref{b}), e.g., they are pseudodistances too. For
example, the above considered functions $\phi (t)=(t-1)^{2}$ and $\phi
(t)=t\ln t$\ lead in this case to the classical Pearson divergence
\begin{equation}
D_{\phi }(p,q)=\sum_{i=1}^{d}\frac{\left( p_{i}-q_{i}\right) ^{2}}{q_{i}}
\label{p}
\end{equation}
and the above mentioned Kullback divergence $D_{\phi }(p,q)\equiv B_{\phi
}(p,q)$ which are asymmetric in $p,q$\ and contradict the triangle
inequality. On the other hand, $\phi (t)=|t-1|$ leads to the L$_{1}$-norm $
||p-q||$\ which is a metric distance and $\phi (t)=(t-1)^{2}/(t+1)$ defines
the LeCam divergence
\begin{equation*}
D_{\phi }(p,q)=\sum_{i=1}^{d}\frac{\left( p_{i}-q_{i}\right) ^{2}}{
p_{i}+q_{i}}
\end{equation*}
which is a squared metric distance (for more about the metricity of $\phi $
-divergences the reader is referred to Vajda (2009)). \medskip

However, there exist also some sharp differences between these two types of
pseudodistances of distributions. One distinguising property of Bregman 
distances is that their use as loss criterion induces 
the conditional expectation as outcoming unique optimal predictor 
from given data (cf.\ Banerjee at al.\ (2005a)); this is for instance
used in Banerjee et al.\ (2005b) for designing generalizations of the 
$k$\textit{-means algorithm} which deals with the special case
of squared Euclidean error (\ref{e}) (cf.\ the seminal work of Lloyd (1982) reprinting a
Technical Report of Bell Laboratories dated by 1957). 
These features are generally not shared by those of the $\phi $-divergences which are
not Bregman distances, e.g., by the Pearson divergence (\ref{p}).
On the other hand, a distinguishing property of $\phi $-divergences is the \textit{information
processing property}, i.e., the impossibility to increase the value $D_{\phi
}(p,q)$ by 
transformations of the observations distributed by p, q and preservation of this value 
by the statistically sufficient transformations
(Csisz\'{a}r (1967), see in this respect also Liese and Vajda (2006)). This property is not
shared by the Bregman distances which are not $\phi $-divergences. For
example, the distributions $p=(1/2,1/4,1/4)$\ and $q=(1,0,0)$\ are mutually
closer (less discernible) in the Euclidean sense (\ref{e}) than their
reductions $\tilde{p}=(1/2,1/2)$\ and $\tilde{q}=(1,0)$ obtained by merging the second and
third observation outcomes into one.\medskip 

Depending on the need to exploit one or the other of these
distinguished properties, the Bregman distances or Csisz\'{a}r divergences
are preferred, and both of them are widely applied in important areas of
information theory, statistics and computer science, for example in\medskip 

\noindent
\textbf{(Ai)} \ \ \textit{information retrieval} (see, e.g., Do  and  Vetterli
(2002), Hertz at al. (2004)), 

\medskip\noindent 
\textbf{(Aii)} \ \textit{optimal decision} (for \textit{general decision} see,
e.g., Boratynska (1997), Freund et al.\ (1997), Bartlett et al.\ (2006), Vajda   and  Zv\'{a}rov\'{a}
(2007), for \textit{speech processing }see, e.g., Carlson
and  Clements (1991), Veldhuis   and  Klabers (2002), 
for \textit{image processing} see, e.g., Xu   and  Osher (2007), 
Marquina  and  Osher (2008), Scherzer et al.\ (2008)), and

\medskip \noindent
\textbf{(Aiii)} \textit{machine learning} (see, e.g., Laferty (1999), 
Banerjee et al. (2005), Amari (2007), Teboulle (2007), Nock   and 
Nielsen (2009)).

\medskip \noindent
\textbf{(Aiv)} \textit{parallel optimization and computing} (see, e.g., Censor and Zenios
(1997)).

\bigskip

In this context it is obvious the importance of the functionals of
distributions which are simultaneously divergences in both the Csisz\'{a}r
and Bregman sense or, more broadly, of the research of relations between the
Csisz\'{a}r and Bregman divergences. This paper is devoted to this research.
It generalizes the separable Bregman distances (\ref{b}) as well as the $
\phi $-divergences (\ref{5}) by introducing the \textit{scaled Bregman
distances}   which for the discrete setup reduce to
\begin{eqnarray}
B_{\phi }(p,q|m) & = & \sum_{i=1}^{d}\Big[ \phi (p_{i}/m_{i})-\phi
(q_{i}/m_{i})  \notag \\
& & -\phi _{+}^{\prime }(q_{i}/m_{i})(p_{i}/m_{i}-q_{i}/m_{i})
\Big] m_{i}  \label{bm}
\end{eqnarray}
for arbitrary finite scale vectors $m=(m_{1},...,m_{d}),$ convex functions $
\phi $ and right-hand derivatives $\phi _{+}^{\prime }$. Obviously, the
uniform scales $m=(1,...,1)$\ 
  lead to the Bregman distances (\ref
{b}) and the probability distribution scales $m=$ $q$ $=(q_{1},...,q_{d})$ 
lead to the $\phi $-divergences (\ref{5}).   We shall work out further interesting
relations of the $B_{\phi }(p,q|m)$\ distances to the $\phi $-divergences $
D_{\phi }(p,q)$ and $D_{\phi }(p,m)\ $ 
and evaluate explicit formulas for the stochastically scaled Bregman distances in arbitrary
exponential families of distributions, including also
the non-discrete setup.\medskip

Section II defines the $\phi $-divergences $D_{\phi }(P,M)$ of general
probability measures $P$ and arbitrary finite measures $M$ and briefly
reviews their basic properties. Section III introduces scaled Bregman
distances $B_{\phi }(P,Q|M)$\ and investigates their relations to the $\phi $
-divergences $D_{\phi }(P,Q)$ and $D_{\phi }(P,M).$ Section IV studies in
detail the situation where all three measures $P,Q,M$ are from the family of
general exponential distributions.\ Finally, Section V illustrates the
results by investigating concrete examples of $P,Q,M$ from classical
statistical families as well as from a family of important random
processes.

\medskip

\paragraph{Notational conventions}
Throughout the paper, $\mathcalM$ denotes the space of all 
  finite measures on a measurable space $({\mathcal{X}},{\mathcal{
A}})$ and $\mathcalP\subset \mathcalM$ the subspace of all probability
measures. Unless otherwise explicitly stated $P,Q,M\,$are mutually
measure-theoretically equivalent measures on $({\mathcal{X}},{\mathcal{A}})$
dominated by a $\sigma $-finite measure $\lambda $ on $({\mathcal{X}},{
\mathcal{A}})$. Then the densities 
\begin{equation}
p={\frac{\mathrm{d}P}{\mathrm{d}\lambda }},\quad q={\frac{\mathrm{d}Q}{
\mathrm{d}\lambda }}\qquad \mbox{and}\qquad m={\frac{\mathrm{d}M}{\mathrm{d}
\lambda }}  \label{d}
\end{equation}
have a common support which will be identified with $\mathcal{X}$ (i.e., the
densities (\ref{d}) are 
positive on $\mathcal{X}$). Unless otherwise
explicitly stated, it is assumed that 
$P,Q\in \mathcalP$, \, $M\in \mathcalM$ and that $\phi :(0,\infty )\mapsto {\mathbb{R}}$ 
is a continuous and convex function. 
It is known that then the possibly infinite
extension $\phi (0)=\lim_{t\downarrow 0}\phi (t)$ and the right-hand
derivatives $\phi _{+}^{\prime }(t)$ for $t\in \lbrack 0,\infty )$ exist,
and that the adjoint function 
\begin{equation}
\phi ^{\ast }(t)=t\phi (1/t)  \label{star}
\end{equation}
is continuous and convex on $(0,\infty )$ with possibly infinite extension $
\phi ^{\ast }(0)$. We shall assume that $\ \phi (1)\equiv \phi ^{\ast
}(1)=0. $

\section{DIVERGENCES}

For $P\,\in \mathcalP$ and $M\in \mathcalM$ we consider 
\begin{equation}
D_{\phi }(P,M)
= \int_{{\mathcal{X}}}\phi \left( {\frac{p}{m}}\right) 
\mathrm{d}M
=\int_{{\mathcal{X}}}m\,\phi \left( {\frac{p}{m}}\right) 
\mathrm{d}\lambda  \qquad (cf.\ (\ref{d}))
\label{div}
\end{equation}
generated by the same convex functions as considered in the formula (\ref
{5}) for discrete $P$\ and $M$. An important special case is
$D_{\phi }(P,Q)$ with $Q\in\mathcalP$. \medskip

The existence (but possible infinity) of the $\phi $-divergences\ follows
from the bounds 
\begin{equation}
\phi _{+}^{\prime }(1)(p-m)\ \leq \ m\,\phi \left( {\frac{p}{m}}\right) \
\leq \ m\,\phi (0)+p\,\phi ^{\ast }(0)
\label{fo.intbounds}
\end{equation}
on the integrand, leading to the $\phi $-divergence bounds
\begin{equation}
\phi _{+}^{\prime }(1)(1-M(\mathcal{X}))\ \leq \ D_{\phi }(P,M)\ \leq \ M(
\mathcal{X})\,\phi (0)+\phi ^{\ast }(0).  \label{1.b}
\end{equation}
The integrand bounds   (\ref{fo.intbounds}) follow by putting $s=1$\ and $t=p/m$ \ in the
inequality 
\begin{equation}
\phi (s)+\phi _{+}^{\prime }(s)(t-s)\leq \phi (t)\leq \phi (0)+t\phi ^{\ast
}(0), 
\label{1.a}
\end{equation}
where the left-hand side is the well-known support line of\textit{\ }$\phi
(t)$ at $t=s.$ The right-hand inequality is obvious for $\phi (0)=\infty $.
If $\phi (0)<\infty $ then it follows by taking $s\rightarrow \infty $\ in
the inequality
\begin{equation*}
\phi (t)\leq \phi (0)\,+\,t\ \frac{\phi (s)-\phi (0)}{s}\ ,
\end{equation*}
obtained from the Jensen inequality for $\phi (t)$\ situated between $\phi
(0)$ and $\phi (s)$. Since the function $\psi (p,m)=m\phi (p/m)$\ is
homogeneous of order 1 in the sense $\psi (tp,tm)=t\psi (p,m)$ for all $t>0,$\ the
divergences (\ref{div}) do n{ot} depend on the choice of the dominating
measure $\lambda $. \medskip

Notice that $D_{\phi }(P,M)$ might be negative. For probability measures $
P,Q $\ the bounds (\ref{1.b}) take on the form
\begin{equation}
0\leq D_{\phi }(P,Q)\leq \phi (0)+\phi ^{\ast }(0),\medskip  \label{1.c}
\end{equation}
and the equalities are achieved under well-known conditions (cf.\ Liese and
Vajda (1987), (2006)): the left equality holds \textit{if} $P=Q$,\ and the
right one holds \textit{if} $P\perp Q$\ (singularity). Moreover, if $\phi
(t) $\ is strictly convex at $t=1,$\ the first \textit{if} can be replaced
by \textit{iff}, \ and in the case $\phi (0)+\phi ^{\ast }(0)<\infty $\ also
the second \textit{if} can be replaced by \textit{iff}. \bigskip

An alternative to the left-hand inequality in (\ref{1.b}), which extends the
left-hand inequality in (\ref{1.c}) including the conditions for the
equality, is given by the following statement (for a systematic theory of $
\phi $-divergences of finite measures we refer to the recent paper of
Stummer and Vajda (2010)).

\bigskip


\paragraph{Lemma 1}

For every $P\in \mathcalP$, $M\in \mathcalM$ one gets the lower
divergence bound 
\begin{equation}
M({\mathcal{X}})\,\phi \left( {\frac{1}{M({\mathcal{X}})}}\right) \leq
D_{\phi }(P,M)\ ,  \label{V.1a}
\end{equation}
where the equality holds if 
\begin{equation}
p\ =\ {\frac{m}{M({\mathcal{X}})}}\text{ \ \ \ \ }P\text{-a.s.}  \label{V.1b}
\end{equation}
If $D_{\phi }(P,M)<\infty $ and $\phi (t)$ is strictly convex at $t=1/M({
\mathcal{X}}),$ the equality in (\ref{V.1a}) holds if and only if (\ref{V.1b}
) holds.

\bigskip

\paragraph{Proof}
By (\ref{div}) and the definition (\ref{star}) of the convex function 
  $\phi^{\ast }$
\begin{equation*}
D_{\phi }(P,M)=\int_{{\mathcal{X}}}\phi ^{\ast }\left( {\frac{m}{p}}\right) 
\mathrm{d}P.
\end{equation*}
Hence by Jensen's inequality 
\begin{equation}
D_{\phi }(P,M)\geq \phi ^{\ast }\left( \int_{{\mathcal{X}}}{\frac{m}{p}}\,
\mathrm{d}P\right) =\phi ^{\ast }(M({\mathcal{X}}))  \label{V.1c}
\end{equation}
which proves the desired inequality (\ref{V.1a}). Since 
\begin{equation*}
{\frac{m}{p}}\ =\ M({\mathcal{X}})\quad P\mbox{-a.\,s.}
\end{equation*}
is the condition for equality in (\ref{V.1c}), the rest is clear from the
easily verifiable fact that $\phi ^{\ast }(t)$ is strictly convex at $t=s$
if and only if $\phi (t)$ is strictly convex at $t=1/s$. 
\hfill $\square $

\bigskip For some of the representation investigations below, it will also
be useful to take into account that for probability measures $P,Q$\ we get
directly from definition (\ref{div}) the \textquotedblleft skew
symmetry\textquotedblright\ $\phi $-divergence formula 
\begin{equation}
D_{\phi ^{\ast }}(P,Q)=D_{\phi }(Q,P)\ ,  
\notag
\end{equation}
as well as the sufficiency of the condition 
\begin{equation}
\phi (t)-\phi ^{\ast }(t)\ \equiv \ \text{constant}\,\mathbf{\cdot }\,(t-1)
\label{after.1.d}
\end{equation}
for the $\phi $-divergence symmetry
\begin{equation}
D_{\phi }(P,Q)=D_{\phi }(Q,P)\mbox{ \ \ for all }P,Q\mbox{\ }.  \label{1.e}
\end{equation}
Liese and Vajda (1987) proved that under the assumed strict convexity of $
\phi (t)$ at $t=1$ the condition (\ref{after.1.d}) is is not only \textit{
sufficient} but also \textit{necessary} for the symmetry (\ref{1.e}).

\section{ SCALED BREGMAN DISTANCES}

Let us now introduce the basic concept of the current paper, which is a
measure-theoretic version of the Bregman distance (\ref{bm}). In this  
definition it is assumed that $\phi$ is a finite convex function in the domain $t>0,$\
continuously extended to $t=0$. As before, $\phi _{+}^{\prime }(t)$\ denotes
the right-hand derivative which for such $\phi (t)$\ exists and $p,q,m$\ are
the densities defined in (\ref{d}).

\bigskip


\paragraph{Definition 1}
The \textit{Bregman distance} of probability measures $P,\,Q$ \textit{scaled}
by an arbitrary measure $M$ on $(\mathcal{X},\mathcal{A})$\
measure-theoretically equivalent with $P,\,Q$ is defined by the formula
\begin{eqnarray}
& & \hspace{-1.1cm} B_{\phi }\left( P,Q\,|\,M\right) \notag\\[0.15cm]
& &\hspace{-1.1cm} =\int_{{\mathcal{X}}}\left[ \phi \left( {
\frac{p}{m}}\right) -\phi \left( {\frac{q}{m}}\right) -\phi _{+}^{\prime
}\left( {\frac{q}{m}}\right) \left( \frac{p}{m}-\frac{q}{m}\right) \right] 
\mathrm{d}M   \notag \\[-0.15cm]
\label{2} \\[-0.15cm]
& &\hspace{-1.1cm} =\int_{{\mathcal{X}}}\left[ m\phi \left( {\frac{p}{m}}\right) -m\phi
\left( {\frac{q}{m}}\right) -\phi _{+}^{\prime }\left( {\frac{q}{m}}\right)
(p-q)\right] \mathrm{d}\lambda.   
\notag \\[-0.35cm]
&&  \notag
\end{eqnarray}
The convex $\phi $\ under consideration can be interpreted
as a generating function of the distance.


\bigskip

\paragraph{Remarks 1}

(1) By putting $t=p/m$\ and $s=q/m$\ in (\ref{1.a}) we find the argument of the
integral in (\ref{2}) to be nonnegative. Hence the Bregman distance $B_{\phi
}\left( P,Q\,|\,M\right) $\ is well-defined by (\ref{2}) 
and
is always nonnegative (possibly infinite).\\ 
\indent (2) Notice that the integrand in the first (respectively second) integral
of \eqref{2} constitutes a function, say, $\widetilde{\Upsilon}(p,q,m)$ 
(respectively $\Upsilon(p,q,m)$)
which is homogeneous of order $0$ (respectively order $1$), i.e., for all $t>0$
there holds $\widetilde{\Upsilon}(tp,tq,tm)=\widetilde{\Upsilon}(p,q,m)$ 
(respectively $\Upsilon(tp,tq,tm)=t\cdot \Upsilon(p,q,m)$).
Analogously, as already partially indicated above, 
the integrand in the first (respectively second) integral
of \eqref{div} is also a function, say, $\widetilde{\psi}(p,m)$ (respectively $\psi(p,m)$)
which is homogeneous of order $0$ (respectively order $1$).\\
\indent (3) In our \textit{measure-theoretic} context \eqref{2} we have incorporated the possible 
non-differentiability of $\phi$ by using its right-hand derivative, which will be
essential at several places below. For general \textit{Banach spaces}, one typically employs
various directional derivatives -- see, e.g., Butnariu and Resmerita (2006) in connection
with different types of convexity properties.
\bigskip

The special scaled Bregman distances $B_{\phi }\left( P,Q\,|\,M\right) $ for
probability scales $M\in \mathcalP$ were introduced by Stummer (2007). 
Let us mention some other important previously considered special cases.

\medskip \noindent \textbf{(a)} For $\mathcal{X}$ finite or countable and
counting   measure $M=\lambda $ some authors were already 
cited above in connection with the formula (\ref{b}) and 
the research areas \textbf{(Ai)} - \textbf{(Aiii)}. 
In addition to them, one can mention also
Byrne (1999),  
Collins et al.\ (2002), 
Murata et al.\ (2004), Cesa-Bianchi and Lugosi (2006). 

\medskip \noindent \textbf{(b)} For open Euclidean set $\mathcal{X}$ and Lebesgue measure $
M=\lambda $ on it one can mention 
Jones and Byrne
(1990),   as well as Resmerita and Anderssen (2007). 
\bigskip

In the rest of this paper, we restrict 
ourselves to the Bregman distances 
$B_{\phi }\left( P,Q\,|\,M\right) $\ scaled by finite measures $M\in \mathcal{
M}$ and to the same class of convex functions as considered in the $\phi $-divergence
formulas (\ref{5}) and (\ref{div}). By using the remark after Definition 1
and applying (\ref{1.a}) we get 
\begin{equation}
D_{\phi }(P,M)\geq D_{\phi }(Q,M)+\int_{\mathcal{X}}\phi _{+}^{\prime
}\left( {\frac{q}{m}}\right) (p-q)\mathrm{d}\lambda  
\notag
\end{equation}
if at least one of the right-hand side expressions is finite. Similarly,
\begin{equation}
B_{\phi }\left( P,Q\,|\,M\right) =D_{\phi }(P,M)-D_{\phi }(Q,M)-\int_{
\mathcal{X}}\phi _{+}^{\prime }\left( {\frac{q}{m}}\right) \mathrm{d}\lambda
\label{1.1}
\end{equation}
if at least two of the right-hand side expressions are finite (which can be
checked, e.g., by using (\ref{1.b}) or (\ref{V.1a})).\medskip

The formula (\ref{2}) simplifies in the important special cases $M=P$\ and $
M=Q$. In the first case, due to $\phi (1)=0$ it reduces to
\begin{eqnarray}
& & \hspace{-1.2cm} B_{\phi }\left( P,Q\,|\,P\right) = \int_{{\mathcal{X}}}\left[ \phi
_{+}^{\prime }\left( {\frac{q}{p}}\right) (q-p)-p\phi \left( {\frac{q}{p}}
\right) \right] \mathrm{d}\lambda \medskip  
\notag \\
& &\hspace{-1.2cm}  = \ \int_{{\mathcal{X}}}\phi _{+}^{\prime }\left( {\frac{q}{p}}\right) (q-p)
\mathrm{d}\lambda -D_{\phi }(Q,P)\ ,  \label{2.bb}
\end{eqnarray}
where the difference (\ref{2.bb}) is meaningful if and only if $D_{\phi
}(Q,P)\equiv D_{\phi ^{\ast }}(P,Q)$ is finite. The nonnegative divergence
measure $\mathcal{B}_{\phi }\left( P,Q\right) :=B_{\phi }\left(
P,Q\,|\,P\right) $ is thus the difference between the nonnegative
dissimilarity measure 
\begin{equation*}
\mathcal{D}_{\phi }\left( Q,P\right) =\int_{{\mathcal{X}}}\phi _{+}^{\prime
}\left( {\frac{q}{p}}\right) (q-p) \, \mathrm{d}\lambda \ \geq \ D_{\phi }(Q,P)
\end{equation*}
and the nonnegative $\phi -$divergence $D_{\phi }(Q,P)$. Furthermore, in the
second special case $M=Q$ the formula (\ref{2}) leads to the equality 
\begin{equation}
B_{\phi }\left( P,Q\,|\,Q\right) =D_{\phi }(P,Q)  \label{2.c}
\end{equation}
without any restriction on $P,Q\in \mathcalP$ as realized already by
Stummer (2007).\bigskip

\paragraph{Conclusion 1}

Equality (\ref{2.c}) -- together with   
the fact that   
$B_{\phi }\left( P,Q\,|\,M\right)$ depends
in general on $M$ (see, e.g., Subsection B below) -- 
shows that the concept of scaled 
Bregman distance (\ref{2}) strictly generalizes the concept of $\phi -$divergence $D_{\phi }(P,Q)$\ of
probability measures $P,Q$.

\bigskip

\paragraph{Example 1}

As an illustration not considered earlier we can take the non-differentiable
function $\phi (t)=|t-1|$\ for which 
\begin{equation*}
B_{\phi }\left( P,Q\,|\,Q\right) =V(P,Q)
\end{equation*}
i.e.,  this particular scaled Bregman distance reduces to the well known total variation.

\bigskip

As demonstrated by an example in the Introduction, measurable
transformations (statistics)
\begin{equation}
T:(\mathcal{X},\mathcal{A})\mapsto (\mathcal{Y},\mathcal{B})  \label{t}
\end{equation}
which are \textit{not} sufficient for the pair $\{P,Q\}$ can increase 
those of the scaled Bregman
distances $B_{\phi }\left( P,Q\,|\,M\right) $\ which are not $\phi $
-divergences. 
On the other hand, the transformations (\ref{t}) which \textit{are}
sufficient for the pair $\{P,Q\}$ need not preserve these distances either. 
Next we formulate conditions under which the scaled Bregman
distances $B_{\phi }\left( P,Q\,|\,M\right) $\ are preserved by
transformations of observations.

\bigskip

\paragraph{Definition 2}

We say that the transformation (\ref{t}) is sufficient for the triplet $
\{P,\,Q,\,M\}$ if there exist measurable functions $g_{P},g_{Q},g_{M}:\mathcal{
Y}\mapsto \mathbb{R}$\ and $h:\mathcal{X}\mapsto \mathbb{R}$\ such that
\begin{eqnarray}
& &p(x)=g_{P}(Tx)h(x),\text{ \ }q(x)=g_{Q}(Tx)h(x)  \notag \\
& &\text{and \ } m(x)=g_{M}(Tx)h(x).  \label{suf}
\end{eqnarray}

If $M$\ is probability measure then our definition reduces to the classical
statistical sufficiency of the statistic $T$\ for the family $\{P,\,Q,\,M\}$
(see pp.\ 18-19 in Lehman (2005)). All transformations (\ref{t}) induce the
probability measures $PT^{-1}$, $QT^{-1}$\ and the finite measure $MT^{-1}$\
on $(\mathcal{Y},\mathcal{B})$. We prove that the scaled Bregman distances
of induced probability measures $PT^{-1}$, $QT^{-1}$\ scaled by $MT^{-1}$\
are preserved by sufficient transformations $T$.

\bigskip
\paragraph{Theorem 1}

The transformations (\ref{t}) sufficient for the triplet $\{P,Q,\,M\}$
preserve the scaled Bregman distances in the sense that
\begin{equation}
B_{\phi }\left( PT^{-1},QT^{-1}\,|\,MT^{-1}\right) =B_{\phi }\left(
P,Q\,|\,M\right) .  \label{eq}
\end{equation}

\paragraph{Proof.}

By (\ref{2}) and (\ref{suf}), the right-hand side of (\ref{eq}) is equal to
\begin{equation}
\int_{\mathcal{X}}\left[ \phi _{P,M}\left( Tx\right) -\phi _{Q,M}\left(
Tx\right) -\Delta _{P,Q,M}\left( Tx\right) \right] \mathrm{d}M  \label{i}
\end{equation}
for
\begin{equation}
\phi _{P,M}\left( y\right) =\phi \left( \frac{g_{P}(y)}{g_{M}(y)}\right) ,
\text{ \ }\phi _{Q,M}\left( y\right) =\phi \left( \frac{g_{Q}(y)}{g_{M}(y)}
\right)  \label{iii}
\end{equation}
and 
\begin{equation}
\Delta _{P,Q,M}\left( y\right) =\phi _{+}^{^{\prime }}\hspace{-0.1cm}\left( \frac{g_{Q}(y)}{
g_{M}(y)}\right) \left( g_{P}(y)-g_{Q}(y)\right) .  \label{ii}
\end{equation}
By Theorem D in Section 39 of Halmos (1964), the integral (\ref{i}) is equal
to
\begin{equation}
\int_{\mathcal{Y}}\left[ \phi _{P,M}\left( y\right) -\phi _{Q,M}\left(
y\right) -\Delta _{P,Q,M}\left( y\right) \right] \mathrm{d}MT^{-1}
\label{iiii}
\end{equation}
and, moreover,
\begin{equation*}
P(T^{-1}B)=\int_{B}g_{P}(y)\, h(T^{-1}y) \, \mathrm{d}\lambda T^{-1}
\end{equation*}
and similarly for $Q$\ instead of $P$. Therefore
\begin{equation*}
\frac{\mathrm{d}PT^{-1}}{\mathrm{d}\lambda T^{-1}}=g_{P}(y)\, h(T^{-1}y)\text{ \ 
and \ }\frac{\mathrm{d}QT^{-1}}{\mathrm{d}\lambda T^{-1}}=g_{Q}(y)\, h(T^{-1}y)
\end{equation*}
which together with\ (\ref{iii}), (\ref{ii}) and (\ref{2}) implies that the
integral (\ref{iiii}) is nothing but the left-hand side of (\ref{eq}). This
completes the proof.\hfill $\square $

\bigskip

\paragraph{Remark 2}
Notice that by means of Remark 1(2) after Definition 1, the assertion of Theorem 1
can be principally related to the preservation of $\phi-$divergences by transformations
which are sufficient for the pair $\{P,Q\}$ .

\bigskip In the rest of this section we discuss some important special classes of
scaled Bregman distances obtained for special distance-generating functions $\phi $.


\subsection{Bregman logarithmic distance}

Let us consider the special function $\phi (t)=t\ln t$. Then $\phi ^{\prime
}(t)=\ln t+1$ so that (\ref{2}) implies 
\begin{eqnarray}
& & B_{t\ln t}\left( P,Q\,|\,M\right) \notag\\
& & =\int_{{\mathcal{X}}}\left[ p\ln {\frac{p
}{m}}-q\ln {\frac{q}{m}}-\left( \ln {\frac{q}{m}}+1\right) (p-q)\right] 
\mathrm{d}\lambda  \notag \\
& &  \notag \\
& &=\int_{{\mathcal{X}}}\left[ p\ln {\frac{p}{m}}-p\ln {\frac{q}{m}}\right] 
\mathrm{d}\lambda  \notag \\
& &  \notag \\
& &=\ \int_{{\mathcal{X}}}p\ln {\frac{p}{q}}\,\mathrm{d}\lambda \ =\ D_{t\ln
t}\left( P,Q\right) \ .  \label{new.1.44}
\end{eqnarray}
Thus, for $\phi (t)=t\ln t$ the Bregman distance $B_{\phi }\left(
P,Q\,|\,M\right)$ exceptionally does not depend on the choice of the scaling and 
reference measures $M$ and $\lambda $; in fact, it always leads to the Kulllback-Leibler
information divergence (relative entropy) $D_{t\ln t}(P,Q)$ (cf.\ Stummer
(2007)). As a side effect, this independence gives also rise to examples for
the conclusion that the validity of \eqref{eq} does generally not imply that
$T$ is sufficient for the triplet $\{P,\,Q,\,M\}$. 


\subsection{Bregman reversed logarithmic distance}

Let now $\phi (t)=-\ln t$ so that $\phi ^{\prime }(t)=-1/t$. Then (\ref{2})
implies 
\begin{eqnarray}
& &\hspace{-1.4cm} B_{-\ln t}\left( P,Q\,|\,M\right) \notag \\
& &\hspace{-1.4cm} =\int_{{\mathcal{X}}}\left[ m\ln {\frac{m
}{p}}-m\ln {\frac{m}{q}}+{\frac{m}{q}}(p-q)\right] \mathrm{d}\lambda \medskip
\label{1.6} \\
& &\hspace{-1.4cm} =D_{t\ln t}(M,P)-D_{t\ln t}(M,Q)+\int_{{\mathcal{X}}}{\frac{mp}{q}}\,
\mathrm{d}\lambda -M({\mathcal{X}})\medskip  \label{1.66} \\
& &\hspace{-1.4cm} =D_{-\ln t}(P,M)-D_{-\ln t}(Q,M)+\int_{{\mathcal{X}}}{\frac{mp}{q}}
\mathrm{d}\lambda -M({\mathcal{X}})  \label{1.666}
\end{eqnarray}
where the equalities (\ref{1.66}) and (\ref{1.666}) hold if at least two out
of the first three expressions on the right-hand side are finite. In
particular, (\ref{1.6}) implies (consistent with (\ref{2.c})) 
\begin{equation}
B_{-\ln t}\left( P,Q\,|\,Q\right) =D_{-\ln t}(P,Q)\quad  \label{1.7}
\end{equation}
and (\ref{1.66}) implies for $D_{t\ln t}(P,Q)<\infty $\ (consistent with
(\ref{2.bb})) 
\begin{equation}
B_{-\ln t}\left( P,Q\,|\,P\right) =\chi ^{2}(P,Q)-D_{t\ln t}(P,Q)\ 
\label{1.8}
\end{equation}
where 
\begin{equation*}
\chi ^{2}(P,Q)=\int_{{\mathcal{X}}}{\frac{(p-q)^{2}}{q}}\,\mathrm{d}\lambda
\end{equation*}
is the well-known Pearson information divergence. From (\ref{1.7}) and (\ref
{1.8}) one can also see that the Bregman distance $B_{\phi }\left(
P,Q\,|\,M\right) $ does in general depend on the choice of the reference
measure $M$.

\subsection{Bregman power distances}

In this subsection we restrict ourselves for simplicity to probability
measures $M\in \mathcalP$, i.e., we suppose $M(\mathcal{X})=1$. Under this
assumption we investigate the   scaled Bregman distances 
\begin{equation}
B_{\alpha }\left( P,Q\,|\,M\right) =B_{\phi _{\alpha }}\left(
P,Q\,|\,M\right) \ ,\quad \alpha \in \mathbb{R},\ \alpha \neq 0,\
\alpha \neq 1  \label{1.9}
\end{equation}
for the family of power convex functions 
\begin{equation}
\phi (t)\equiv \phi _{\alpha }(t)={\frac{t^{\alpha }-1}{\alpha (\alpha -1)}}
\mbox{ \ \ with \ }\phi _{\alpha }^{\prime }(t)={\frac{t^{\alpha -1}}{\alpha
-1}\ .}  \label{1.10}
\end{equation}

\noindent For comparison and representation purposes, we use for 
  $P$ (and analogously for $Q$ instead of $P$) the power divergences
\begin{eqnarray}
& & \hspace{-0.9cm} D_{\alpha }(P,M) = D_{\phi _{\alpha }}(P,M) \notag\\[0.2cm]
& & \hspace{-0.9cm} =\frac{1}{\alpha (\alpha -1)}
\left[ \int_{{\mathcal{X}}}p^{\alpha }\,m^{1-\alpha }\,\mathrm{d}\lambda -1
\right] \bigskip  
\notag \\[0.2cm]
& &\hspace{-0.9cm} =\frac{\exp \rho _{\alpha }(P,M)-1}{\alpha (\alpha -1)}\ \text{\ \ with 
}\rho _{\alpha }(P,M)=\ln \int_{{\mathcal{X}}}p^{\alpha }\,m^{1-\alpha }\,
\mathrm{d}\lambda  \notag \\
\label{101}
\end{eqnarray}
of real powers $\alpha $\ different from $0$\ and $1$, studied for arbitrary
probability measures $P,M$\ in Liese and Vajda (1987). They are one-one
related to the R\'{e}nyi divergences
\begin{equation*}
R_{\alpha }(P,M)=\frac{\rho _{\alpha }(P,M)}{\alpha (\alpha -1)},\mbox{ \ \ }
\alpha \in \mathbb{R},\mbox{ }\alpha \neq 0,\mbox{ }\alpha \neq 1,
\end{equation*}
introduced in Liese and Vajda (1987) as an extension of the original
narrower class of the divergences 
\begin{equation*}
R_{\alpha }(P,M)=\frac{\rho _{\alpha }(P,M)}{\alpha -1},\mbox{ \ \
}\alpha >0,\mbox{ }\alpha \neq 1
\end{equation*}
of R\'{e}nyi (1961).\bigskip

Returning now to the Bregman power distances,  
observe that if $D_{\alpha }(P,M)+D_{\alpha }(Q,M)\ $is finite then (\ref
{1.1}), (\ref{1.9}) and (\ref{1.10}) imply for $\alpha \neq 0,\ \alpha \neq
1 $ 
\begin{eqnarray}
& &\hspace{-0.4cm}
B_{\alpha }(P,Q\,|\,M) \notag \\ 
& &\hspace{-0.4cm} = -D_{\alpha }(Q,M)-{
\frac{1}{\alpha -1}}\int_{{\mathcal{X}}}\left( {\frac{q}{m}}\right) ^{\alpha
-1}(p-q)\,\mathrm{d}\lambda \medskip  
\notag \\[0.15cm] 
& &\hspace{-0.4cm} = D_{\alpha }(P,M)-D_{\alpha }(Q,M) \notag\\[0.1cm]
& & \hspace{0.7cm} - {\frac{1}{\alpha -1}}
\int_{{\mathcal{X}}}\left[ \left( {\frac{q}{m}}\right) ^{\alpha
-1}\!\!\!p-\left( {\frac{q}{m}}\right) ^{\alpha }m\right] \mathrm{d}\lambda
\medskip  
\notag \\[0.15cm]
& &\hspace{-0.4cm} = D_{\alpha }(P,M)-(1\!-\!\alpha )\,D_{\alpha }(Q,M) \notag \\[0.1cm]
& & \hspace{0.7cm} -{\frac{1
}{\alpha \!-\!1}}\left[ \int_{{\mathcal{X}}}\left( \frac{q}{m}\right)
^{\alpha -1}\!\!\!p\,\mathrm{d}\lambda -1\right] .  \label{1.11b}
\end{eqnarray}
In particular, we get from here (consistent with (\ref{2.c})) 
\begin{equation}
B_{\alpha }(P,Q\,|\,Q)=D_{\alpha }(P,Q)\quad  
\notag
\end{equation}
and in case of $D_{\alpha }(Q,P)\equiv D_{1-\alpha }(P,Q)<\infty $ also 
\begin{eqnarray}
&& \hspace{-1.1cm} B_{\alpha }(P,Q\,|\,P) = (\alpha -2)\,D_{\alpha -1}(Q,P)+(\alpha
-1)\,D_{\alpha }(Q,P)  \notag \\[-0.15cm]
\notag \\[0.1cm]
&& 
\equiv (\alpha -2)\,D_{2-\alpha }(P,Q)+(\alpha -1)\,D_{1-\alpha }(P,Q).
\notag 
\end{eqnarray}

In the following theorem, and elsewhere in the sequel, we use the simplified
notation
\begin{equation}
D_{1}(P,M)=D_{t\ln t}(P,M)\mbox{ \ and \ }D_{0}(P,M)=D_{-\ln t}(P,M)
\notag 
\end{equation}
for the probability measures $P,M$\ under consideration (and also later on
where $M$ is only a finite measure). This step is motivated by the limit
relations 
\begin{eqnarray}
\lim_{\alpha \downarrow 0}D_{\alpha }(P,M)&=& D_{-\ln t}(P,M)\mbox{ \quad and \ } \notag \\
\lim_{\alpha \uparrow 1}D_{\alpha }(P,M)&=& D_{t\ln t}(P,M)  \label{1.101}
\end{eqnarray}
proved as Proposition 2.9 in Liese and Vajda (1987) for arbitrary
probability measures $P,M$. Applying these relations to the Bregman
distances, we obtain


\bigskip

\paragraph{Theorem 2}

If $D_{0}(P,M)+D_{0}(Q,M)<\infty $ then 
\begin{eqnarray}
& &\lim_{\alpha \downarrow 0}B_{\alpha }(P,Q\,|\,M) \notag \\
&& =D_{0}(P,M)-D_{0}(Q,M)+\int_{{\mathcal{X}}}{\frac{mp}{q}}\,\mathrm{d}
\lambda -1   \label{1.14} \\
& & =  B_{-\ln t}(P,Q\,|\,M).  \label{1.144}
\end{eqnarray}
If $D_{1}(P,M)+D_{1}(Q,M)<\infty $ and 
\begin{eqnarray}
& & \lim_{\beta \downarrow 0}\int_{{\mathcal{X}}}{\frac{(q/m)^{-\beta }-1}{\beta 
}}\,\mathrm{d}P \notag \\
& &= \int_{{\mathcal{X}}}\lim_{\beta \downarrow 0}{\frac{
(q/m)^{-\beta }-1}{\beta }}\,\mathrm{d}P  
= -\int_{{\mathcal{X}}}\ln {\frac{q}{m}}\,\mathrm{d}P  
\label{1.145}
\end{eqnarray}
then 
\begin{eqnarray}
& & \lim_{\alpha \uparrow 1}B_{\alpha }(P,Q\,|\,M) = D_{1}(P,M)-\int_{{\mathcal{
X}}}\ln {\frac{q}{m}}\,\mathrm{d}P  \label{1.15} \\[0.2cm]
& &= D_{1}(P,Q)\ =\ B_{t\ln t}(P,Q\,|\,M)\ .  \label{1.155}
\end{eqnarray}

\bigskip

\paragraph{Proof}

If $0<\alpha <1$\ then $D_{\alpha }(P,M),$\ $D_{\alpha }(Q,M)$\ are finite
so that (\ref{1.11b}) holds. Applying the first relation of (\ref{1.101}) in
(\ref{1.11b})\ we get (\ref{1.14}) where the right hand side is well defined
because  $D_{0}(P,M)$ $+\ D_{0}(Q,M)$\ is by assumption finite.
Similarly, by using the second relation of (\ref{1.101}) and the assumption (
\ref{1.145}) in (\ref{1.11b})\ we end up at (\ref{1.15}) where the
right-hand side is well defined because $D_{1}(P,M$)$+D_{1}(Q,M$)\ is
assumed to be finite. The identity (\ref{1.144}) follows from (\ref{1.14}), (
\ref{1.666}) and the identity (\ref{1.155}) from (\ref{1.15}), (\ref
{new.1.44}).\hfill $\square $

\bigskip

Motivated by 
this theorem, we introduce for all probability measures $P,\,Q,\,M$ under
consideration the simplified notations 
\begin{equation}
B_{1}(P,Q\,|\,M)\ =\ B_{t\ln t}(P,Q\,|\,M)  \label{new.1.16b}
\end{equation}
and 
\begin{equation}
B_{0}(P,Q\,|\,M)\ =\ B_{-\ln t}(P,Q\,|\,M)\ ,  \label{new.1.16c}
\end{equation}
and thus, (\ref{1.155}) and (\ref{1.144}) become 
\begin{equation}
B_{1}(P,Q\,|\,M)=\lim_{\alpha \uparrow 1}B_{\alpha }(P,Q\,|\,M)
\notag
\end{equation}
and
\begin{equation}
B_{0}(P,Q\,|\,M)=\lim_{\alpha \downarrow 0}B_{\alpha }(P,Q\,|\,M).
\notag
\end{equation}
Furthermore, in these notations the relations 
(\ref{new.1.44}), (\ref{1.7}) and (\ref{1.8}) reformulate (under the
corresponding assumptions) as follows 
\begin{equation}
B_{1}(P,Q\,|\,M)\ =\ D_{1}(P,Q)\ ,  
\notag
\end{equation}
\begin{equation}
B_{0}(P,Q\,|\,Q)=D_{0}(P,Q)  
\notag
\end{equation}
and
\begin{eqnarray}
B_{0}(P,Q\,|\,P) &=&\chi ^{2}(P,Q)-D_{1}(P,Q)  \notag \\[0.1cm]
&=&2\,D_{2}(P,Q)-D_{1}(P,Q).  \label{new.1.19}
\end{eqnarray}

\bigskip

\paragraph{Remark 3}

The power divergences $D_{\alpha }(P,Q)$\ are usually applied in the
statistics as criteria of discrimination or goodness-of-fit between the
distributions $P$\ and $Q$. The scaled Bregman distances $B_{\alpha
}(P,Q\,|\,M)\ $as generalizations of the power divergences $D_{\alpha
}(P,Q)\equiv B_{\alpha }(P,Q\,|\,Q)$\ allow 
to extend the 2D-discrimination
plots $\left\{ \left[ D_{\alpha }(P,Q);\text{ }\alpha \right] :c\leq \alpha
\leq d\right\} \subset \mathbb{R}^{2}$ into more informative 3D\textit{
-discrimination plots} 
\begin{equation}
\left\{ \left[ B_{\alpha }(P,Q\text{ }|\,\beta P+(1-\beta )Q);\text{ }\alpha
;\text{ }\beta \right] :c\leq \alpha \text{, }\beta \leq d\right\} \subset 
\mathbb{R}^{3}  \label{dplot}
\end{equation}
reducing to the former ones for $\beta =0$. 
The simpler 2D-plots known under the name $Q$--$Q$-plots are famous tools
for the exploratory data analysis. It is easy to consider that the computer-aided
appropriately coloured projections of the 3D-plots (\ref{dplot}) allow much more
intimate insight into the relation between data and their statistical models.
Therefore this computer-aided 3D-exploratory analysis deserves 
a deeper attention and research.
The next example presents 
projections of two such plots obtained for
a binomial model $P$ and its data based binomial alternative $Q$.

\bigskip 

\paragraph{Example 2}
Let $P=\text{Bin}(n,\widetilde{p})$ be a binomial distribution with parameters 
$n$, $\widetilde{p}$ (with a slight abuse of notation), 
and $Q=\text{Bin}(n,\widetilde{q})$. 
Figure 1 presents projections of the corresponding 3D-discrimination plots (\ref{dplot})
for $0.2\leq \alpha \leq 2$\ and $0 \leq \beta \leq 1$,
where the Subfigure (a) used the parameter constellation
$n=10$, $\widetilde{p}=0.25$, $\widetilde{q}=0.20$ whereas the Subfigure (b) used $n=10$, $\widetilde{p}=0.25$, 
$\widetilde{q}=0.30$.
In both cases, the ranges of $B_{\alpha }(P,Q\text{ }|\,\beta P+(1-\beta )Q)$
are subsets of the interval $[0.06,0.088]$.

\begin{figure*}[!t]
\centerline{
\subfigure[$\widetilde{p}=0.25$, $\widetilde{q}=0.20$]{\includegraphics[width=2.5in]{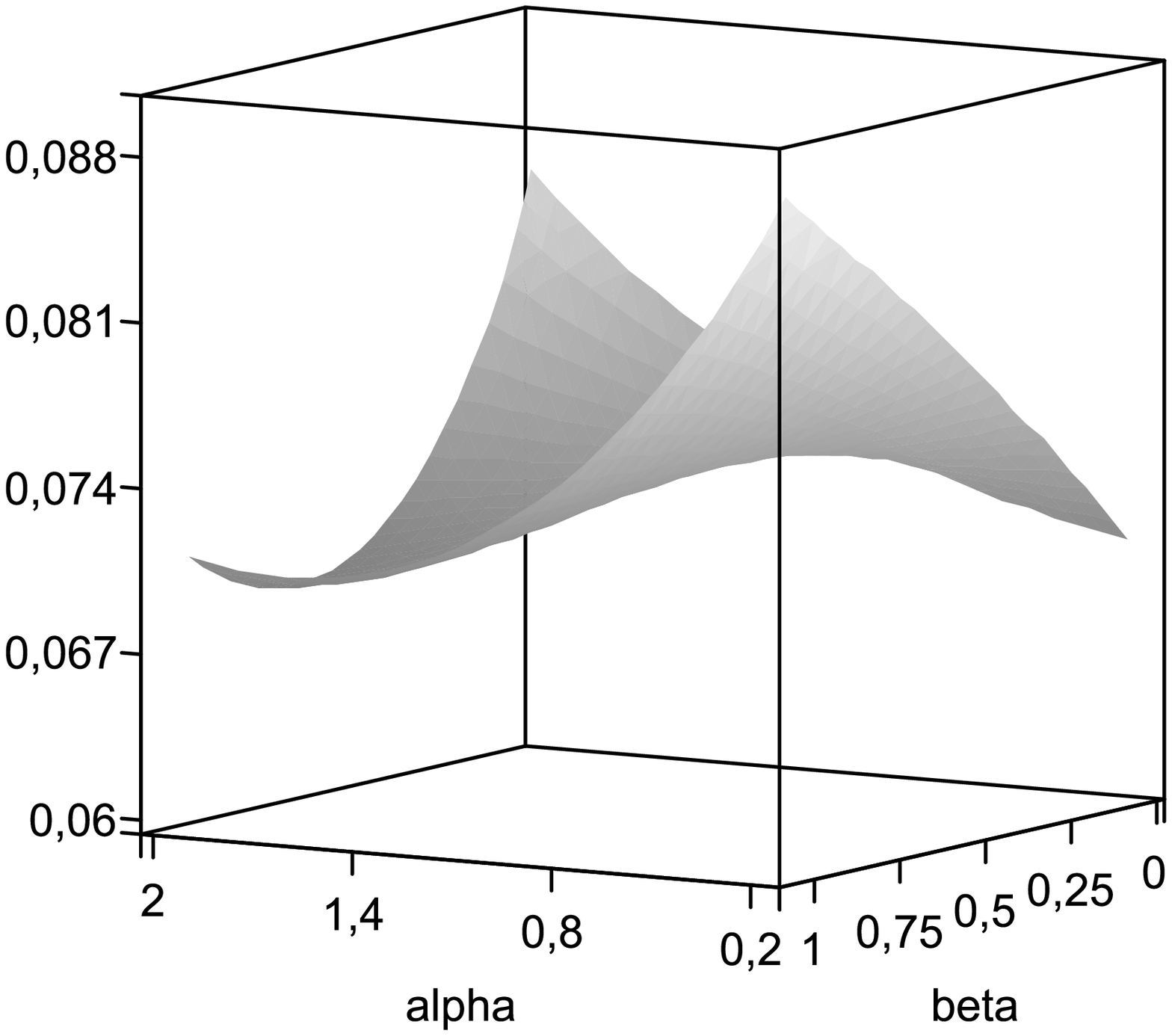}
\label{fig_first_case}}
\hfil
\subfigure[$\widetilde{p}=0.25$, $\widetilde{q}=0.30$]{\includegraphics[width=2.5in]{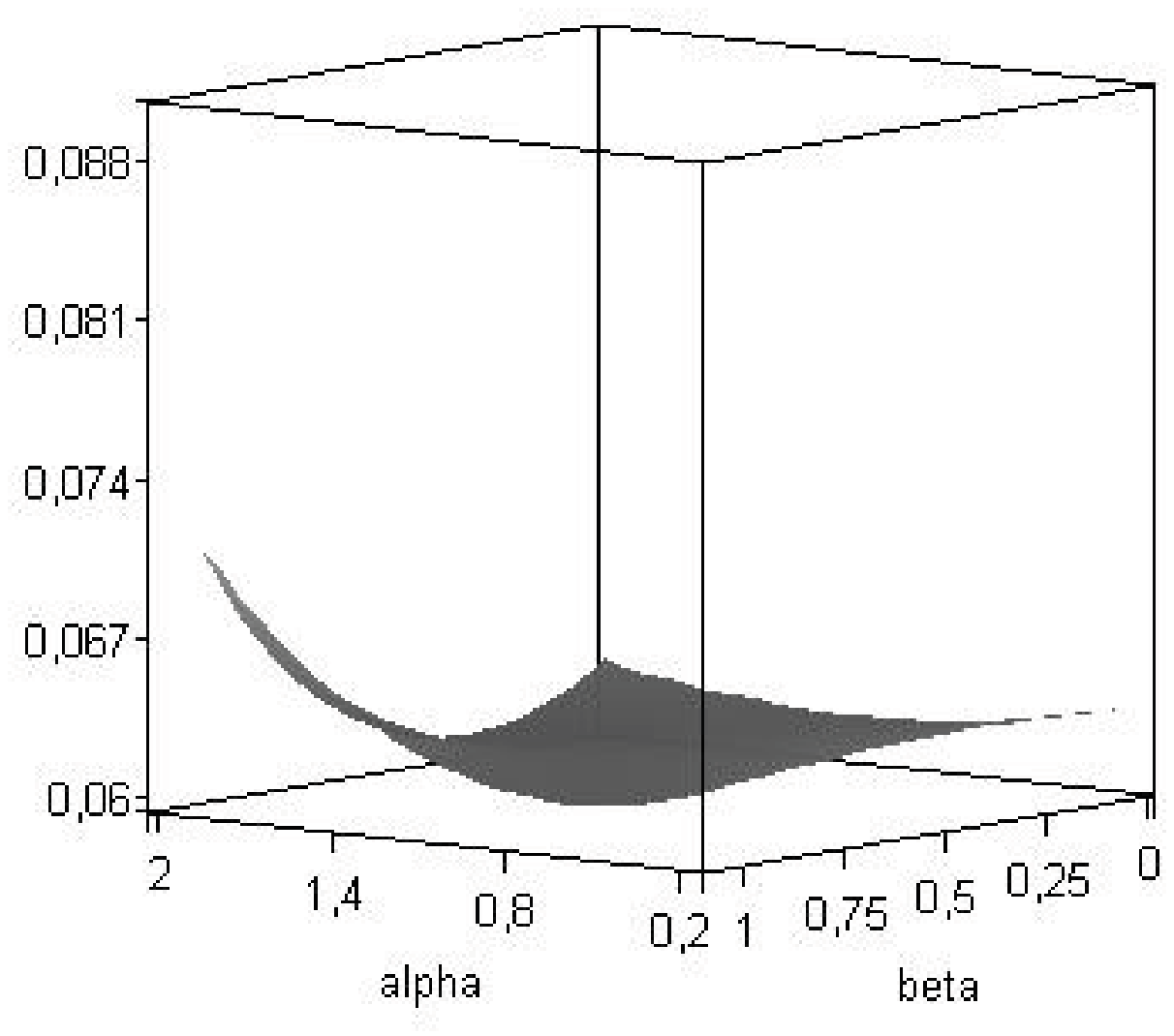}
\label{fig_second_case}}
}
\caption{3D-discrimination plots (\ref{dplot}) for $P=\text{Bin}(10,\widetilde{p})$, 
$Q=\text{Bin}(10,\widetilde{q})$
with $0.2\leq \alpha \leq 2$\ and $0 \leq \beta \leq 1$.}
\label{fig_bregbin}
\end{figure*}

%


\section{EXPONENTIAL FAMILIES}

In this section we show that the scaled Bregman power distances $B_{\alpha
}(P,Q\,|\,M)$ can be \textit{explicitly} \textit{evaluated} for 
probability measures $P,\,Q,\,M$
from exponential families. Let us
restrict ourselves to the Euclidean observation
spaces $(\mathcal{X},\mathcal{A})\subseteq ({\mathbb{R}}^{d},{\mathcal{B}}
^{d})$ and denote by $x\cdot \theta $\ the scalar product of $x,\theta \in {
\mathbb{R}}^{d}$. The convex extended real valued function 
\begin{equation}
b(\theta )=\ln \int_{{\mathbb{R}}^{d}}e^{x\cdot \theta }\mathrm{d}\lambda
(x),\qquad \theta \in {\mathbb{R}}^{d}\, ,  \label{E.2}
\end{equation}
and the\ convex set 
\begin{equation}
\Theta =\{\theta \in {\mathbb{R}}^{d}:b(\theta )<\infty \}  
\notag
\end{equation}
define on $(\mathcal{X},\mathcal{A})$  an \textit{exponential family of
probability measures} $\{P_{\theta }:\theta \in \Theta \}$ with the
densities 
\begin{equation}
p_{\theta }(x)\equiv {\frac{\mathrm{d}P_{\theta }}{\mathrm{d}\lambda }}
(x)=\exp \{x\cdot \theta -b(\theta )\},\quad x\in {\mathbb{R}}^{d},\quad
\theta \in \Theta .  \label{E.1}
\end{equation}
The cumulant function $b(\theta )$ is\ infinitely differentiable
on the interior $\mathring{\Theta}$ with the gradient
\begin{equation}
\bigtriangledown b(\theta )=\left( \frac{\partial }{\partial \theta _{1}}
,...,\frac{\partial }{\partial \theta _{d}}\right) \, b(\theta )\text{, \ \ \ 
}\theta \in \mathring{\Theta}.  
\notag
\end{equation}
Note that (\ref{E.1}) are exponential type densities in the \textit{natural form}.
All exponential 
type distributions such as Poisson, normal etc. can be
transformed to into this form (cf., e.g., Brown (1986)). $\medskip $

The formula 
\begin{equation}
\int_{{\mathbb{R}}^{d}}e^{x\cdot \theta }\ \mathrm{d}\lambda (x)\ =\
e^{b(\theta )},\text{ \ \ \ }\theta \in \Theta  \label{E.4}
\end{equation}
follows from (\ref{E.2}) and implies 
\begin{equation}
\int_{{\mathbb{R}}^{d}}x\,e^{x\cdot \theta }\ \mathrm{d}\lambda (x)\ =\
e^{b(\theta )}\nabla b(\theta ),\text{ \ \ \ }\theta \in \mathring{\Theta}.
\label{E.6}
\end{equation}
Both formulas (\ref{E.4}) and (\ref{E.6}) will be useful in the sequel. 

\medskip 

\noindent We are interested in the scaled Bregman power distances 
\begin{equation}
B_{\alpha }\left( P_{\theta _{1}},P_{\theta _{2}}\,|\,P_{\theta _{0}}\right)
\quad \mbox{for}\ \theta _{0},\,\theta _{1},\,\theta _{2}\in \Theta ,\
\alpha \in {\mathbb{R}}.  
\notag
\end{equation}
Here $P_{\theta _{1}},P_{\theta _{2}},P_{\theta _{0}}$ are
measure-theoretically equivalent probability measures, so that we can turn
attention to the formulas (\ref{1.11b}), (\ref{new.1.44}), (\ref{1.666}),
and (\ref{new.1.16b}) to (\ref{new.1.19}), promising to reduce the
evaluation of $B_{\alpha }(P_{\theta _{1}},P_{\theta _{2}}\,|\,P_{\theta
_{0}})$ to the evaluation of the power divergences $D_{\alpha }(P_{\theta
_{1}},P_{\theta _{2}})$. Therefore we first study these divergences and in
particular verify their finiteness, which was a sufficient condition for the
applicability of the formulas (\ref{1.11b}), (\ref{new.1.44}) and (\ref
{1.666}). To begin with, let us mention the following 
well-established representation: 

\bigskip

\paragraph{Theorem 3}

If $\alpha \in {\mathbb{R}}$ differs from $0$ and $1$, then 
the power divergence $D_{\alpha}(P_{\theta _{1}},P_{\theta _{2}})$ is 
for all $\theta _{1},\,\theta_{2}\in \Theta$\,  finite and given by the expression
\begin{eqnarray}
& &\hspace{-0.8cm}  
{\frac{\exp {
\big\{} b(\alpha \theta _{1}+(1-\alpha )\,\theta _{2})-\alpha b(\theta
_{1})-(1-\alpha )\,b(\theta _{2}) {\big\}}-1 }{\alpha (\alpha -1)} \ .} \notag \\[-0.1cm]
\label{E.8}
\end{eqnarray}

\vspace{-0.3cm}\noindent
In particluar, it is invariant with respect to the shifts
of the cumulant function linear in $\theta \in \Theta$ in the sense
that it coincides with the power divergence
$D_{\alpha}(\tilde{P}_{\theta _{1}},\tilde{P}_{\theta _{2}})$
in the exponential family with the cumulant function 
$\tilde{b}(\theta) = b(\theta) + c + v\cdot \theta$ 
where $c$ is a real number and $v$ a $d-$vector.

\bigskip

This can be easily seen by slightly extending (\ref{101}) to get for arbitrary $\alpha \in {\mathbb{
R}}$ and $\theta _{1},\,\theta _{2}\in \Theta $ 
\begin{eqnarray}
& &\hspace{-0.7cm}  1+\alpha \cdot (\alpha-1) \cdot D_{\alpha}(P_{\theta _{1}},P_{\theta _{2}}) 
 = 
\int_{\mathbb{R}
^{d}}p_{\theta _{1}}^{\alpha }\, p_{\theta _{2}}^{1-\alpha }\ \mathrm{d}\lambda
\notag
\\
& &\hspace{-0.7cm} = 
{\frac{\int_{\mathbb{R}^{d}}\exp \{x\cdot \lbrack \alpha \theta
_{1}+(1-\alpha )\,\theta _{2}]\}\ \mathrm{d}\lambda (x)}{\exp \{\alpha
b(\theta _{1})+(1-\alpha )\,b(\theta _{2})\}}}  \notag 
\end{eqnarray}
which together with \eqref{E.4} gives the desired result.

\bigskip \noindent The skew symmetry as well as the remaining power divergences $D_0(P_{\theta_1},P_{
\theta_2})$ and $D_1(P_{\theta_1},P_{\theta_2})$ are 
given in the next, straightforward theorem.


\bigskip

\paragraph{Theorem 4}

For all $\theta _{1},\,\theta _{2}\in \Theta $ and $\alpha \in {\mathbb{R}}$
different from 0 and 1   there holds
\begin{equation}
D_{\alpha }\left( P_{\theta _{2}},P_{\theta _{1}}\right) \ =\ D_{1-\alpha
}\left( P_{\theta _{1}},P_{\theta _{2}}\right)  
\notag
\end{equation}
and for $\theta _{2}\in \mathring{\Theta}$ 
\begin{eqnarray}
&&\hspace{-1.35cm} D_{-\ln t}\left( P_{\theta _{1}},P_{\theta _{2}}\right)  = D_{0}\left(
P_{\theta _{1}},P_{\theta _{2}}\right)  =  \lim_{\alpha \downarrow
0}D_{\alpha }\left( P_{\theta _{1}},P_{\theta _{2}}\right)  
\notag \\[0.15cm]
& &\hspace{-1.35cm} = b(\theta _{1})-b(\theta _{2})-\nabla b(\theta _{2})\,(\theta
_{1}-\theta _{2})  \label{E.14} \\[0.15cm]
& &\hspace{-1.35cm} = \lim_{\alpha \uparrow 1}D_{\alpha }\left( P_{\theta _{2}},P_{\theta
_{1}}\right)  = D_{1}\left( P_{\theta _{2}},P_{\theta _{1}}\right)  = 
D_{t\ln t}\left( P_{\theta _{2}},P_{\theta _{1}}\right).  \label{new.E.14b}
\end{eqnarray}

\bigskip  

The \textit{main} result of this section is the following representation theorem for
Bregman distances in exponential families. We formulate this in terms of 
the functions 
\begin{equation}
\rho _{\alpha }(\theta _{1},\theta _{2})=b\Big(\alpha \theta _{1}+(1-\alpha
)\,\theta _{2}\Big)-\alpha b(\theta _{1})-(1-\alpha )\,b(\theta _{2})
\label{E.10}
\end{equation}
(where the right hand side is finite if $0\leq \alpha \leq 1$),
as well as the functions $\sigma _{\alpha }(\theta_{0},\theta _{1},\theta _{2})$ ($
\alpha \in \mathbb{R}$, $\theta_{0},\theta _{1},\theta _{2}\in \Theta$)
de\-fi\-ned 
as the difference
\begin{equation}
\sigma _{\alpha }(\theta _{0},\,\theta _{1},\,\theta _{2})=\sigma _{\alpha
}^{I}(\theta _{0},\,\theta _{1},\,\theta _{2})-\sigma _{\alpha }^{II}(\theta
_{0},\,\theta _{1},\,\theta _{2})  \label{E.15}
\end{equation}
of the nonnegative (possibly infinite)
\begin{equation}
\sigma _{\alpha }^{I}(\theta _{0},\,\theta _{1},\,\theta _{2})=b{\Big(}
\alpha \,\theta _{1}+(1-\alpha )\,\left[ \theta _{1}-\theta _{2}+\theta _{0}
\right] {\Big)}  \label{E.15a}
\end{equation}
and the finite
\begin{equation}
\sigma _{\alpha }^{II}(\theta _{0},\,\theta _{1},\,\theta _{2})=\alpha
\,b(\theta _{1})+(1-\alpha )\, {\Big[} b(\theta _{1})-b(\theta
_{2})+b(\theta _{0}){\Big]} \ .  \label{E.15b}
\end{equation}

\noindent
Alternatively,
\begin{eqnarray}
& &\hspace{-1.4cm} \sigma _{\alpha }(\theta _{0},\,\theta _{1},\,\theta _{2}) = \rho _{\alpha
}(\theta _{1},\theta _{0}+\theta _{1}-\theta _{2}) \notag \\ 
& &\hspace{-0.9cm} +(1-\alpha )\left[ b(\theta _{0}+\theta _{1}-\,\theta _{2})-b(\theta
_{0})-b(\theta _{1})+b(\theta _{2})\right] .  \label{sigma}
\end{eqnarray}


\paragraph{Theorem 5}

Let $\theta _{0},\,\theta _{1},\,\theta _{2}\in \Theta $ be arbitrary. If $
\alpha (\alpha -1)\neq 0$ then the Bregman distance of the exponential
family distributions $P_{\theta _{1}}$ and $P_{\theta _{2}}$ scaled by $
P_{\theta _{0}}$ is given by the formula 
\begin{eqnarray}
& & \hspace{-1.3cm} B_{\alpha }\left( P_{\theta _{1}},P_{\theta _{2}}\,|\,P_{\theta
_{0}}\right) \notag\\
& & \hspace{-1.3cm}  ={\frac{\exp \rho _{\alpha }(\theta _{1},\theta _{0})}{\alpha
(\alpha -1)}}+{\frac{\exp \rho _{\alpha }(\theta _{2},\theta _{0})}{\alpha }}
+\,{\frac{\exp \sigma _{\alpha }(\theta _{0},\theta _{1},\theta _{2})}{
1-\alpha }}.  \label{E.16}
\end{eqnarray}
If $\theta _{0}$ respectively $\,\theta _{1}$ is from the interior $
\mathring{\Theta}$, \ then the limiting Bregman power distances are 
\begin{eqnarray}
& & \hspace{-1.3cm} B_{0}\left( P_{\theta _{1}},P_{\theta _{2}}\,|\,P_{\theta _{0}}\right) \notag \\
& & 
= \ 
b(\theta _{1})-b(\theta _{2})-\nabla b(\theta _{0})\,(\theta _{1}-\theta
_{2}) \notag \\
& & \hspace{0.45cm} + \exp \sigma _{0}(\theta _{0},\theta _{1},\theta _{2})-1  \label{E.18}
\end{eqnarray}
respectively 
\begin{equation}
B_{1}\left( P_{\theta _{1}},P_{\theta _{2}}\,|\,P_{\theta _{0}}\right)
=b(\theta _{2})-b(\theta _{1})-\nabla b(\theta _{1})\,(\theta _{2}-\theta
_{1}) \ .  \label{E.19}
\end{equation}

\vspace{-0.3cm}\noindent
In particluar, all scaled Bregman distances (\ref{E.16}) - (\ref{E.19})
are invariant with respect to the shifts
of the cumulant function linear in $\theta \in \Theta$ in the sense
that they coincide with the scaled Bregman distances 
$B_{\alpha }\left( \tilde{P}_{\theta _{1}},\tilde{P}_{\theta _{2}}\,|\,\tilde{P}_{\theta
_{0}}\right)$
in the exponential family with the cumulant function 
$\tilde{b}(\theta) = b(\theta) + c + v\cdot \theta$ 
where $c$ is a real number and $v$ a $d-$vector.

\bigskip

\paragraph{Proof}

(a) \ By (\ref{E.1}) it holds for every $\alpha \in {\mathbb{R}}$ and $
\theta _{0},\theta _{1},\,\theta _{2}\in \Theta $ 
\begin{eqnarray*}
&&\left( {\frac{p_{\theta _{2}}(x)}{p_{\theta _{0}}(x)}}\right) ^{\alpha
-1}p_{\theta _{1}}(x) \\[0.1cm]
& &= \ \exp {\Big\{}(\alpha -1)\big[x\cdot (\theta _{2}-\theta _{0})-(b(\theta
_{2})-b(\theta _{0}))\big] \notag \\
& & \hspace{4.75cm} +x\cdot \theta _{1}-b(\theta _{1}){\Big\}} \\[0.1cm]
& &= \  \exp {\Big\{}x\cdot \big(\alpha \,\theta _{1}+(1-\alpha )\,\left[
\theta _{1}-\theta _{2}+\theta _{0}\right] \big) \notag \\
& & \hspace{4.45cm} -\sigma _{\alpha
}^{II}(\theta _{0},\,\theta _{1},\,\theta _{2}){\Big\}}
\end{eqnarray*}
with $\sigma _{\alpha }^{II}(\theta _{0},\,\theta _{1},\,\theta _{2})$\ from
(\ref{E.15b}). Since (\ref{E.4}) leads to 
\begin{eqnarray*}
& & \int_{\mathbb{R}^{d}}\exp \Big\{x\cdot \Big(\alpha \,\theta _{1}+(1-\alpha
)\,\left[ \theta _{1}-\theta _{2}+\theta _{0}\right] \Big)\Big\}\,\mathrm{d}
\lambda \notag \\[0.1cm]
& & = \ \exp \sigma _{\alpha }^{I}(\theta _{0},\,\theta _{1},\,\theta
_{2}) \notag
\end{eqnarray*}
for $\sigma _{\alpha }^{I}(\theta _{0},\,\theta _{1},\,\theta _{2})$\ given
by (\ref{E.15a}), it holds
\begin{equation}
\int_{{\mathcal{X}}}\left( {\frac{p_{\theta _{2}}}{p_{\theta _{0}}}}\right)
^{\alpha -1}p_{\theta _{1}}\,\mathrm{d}\lambda \ =\ \exp \sigma _{\alpha
}(\theta _{0},\theta _{1},\theta _{2})  \label{E.21}
\end{equation}
where $\sigma _{\alpha }(\theta _{0},\theta _{1},\theta _{2})$ was defined
in (\ref{E.15}). Now, by 
plugging
\begin{equation*}
P=P_{\theta _{1}},\quad Q=P_{\theta _{2}},\quad M=P_{\theta _{0}}\text{ \ \
\ (cf. (\ref{E.1}))}
\end{equation*}
in (\ref{1.11b}), we get for $\alpha (\alpha -1)\neq 0$ the Bregman distances 
\begin{eqnarray}
& & B_{\alpha }\left( P_{\theta _{1}},P_{\theta _{2}}\,|\,P_{\theta _{0}}\right) \notag \\[0.1cm]
& &= \ D_{\alpha }\left( P_{\theta _{1}},P_{\theta _{2}}\right) -(1-\alpha
)\,D_{\alpha }\left( P_{\theta _{2}},P_{\theta _{0}}\right)  
\notag \\[0.1cm]
&&\hspace{0.65cm} + \ {\frac{1}{1-\alpha }}\left[ \int_{{\mathcal{X}}}\left( {\frac{
p_{\theta _{2}}}{p_{\theta _{0}}}}\right) ^{\alpha -1}p_{\theta _{1}}\,
\mathrm{d}\lambda -1\right] .  \label{E.20}
\end{eqnarray}
By combining the power divergence formula 
\eqref{E.8} with \eqref{E.10},
one ends up with 
$D_{\alpha }\left( P_{\theta _{1}},P_{\theta _{2}}\right) ={\frac{\exp\{\rho
_{\alpha }(\theta _{1},\theta _{2})\}-1}{\alpha (\alpha -1)}} $
which together with (\ref{E.21})
and (\ref{E.20}) leads to the desired representation (\ref{E.16}).

\medskip

(b) \ By the definition of $B_{0}(P,Q\,|\,M)$ in 
(\ref{new.1.16c}) and
by (\ref{1.14}) 
\begin{eqnarray*}
& & B_{0}\left( P_{\theta _{1}},P_{\theta _{2}}\,|\,P_{\theta _{0}}\right) \notag \\
& &  =\
D_{0}\left( P_{\theta _{1}},P_{\theta _{0}}\right) -D_{0}\left( P_{\theta
_{2}},P_{\theta _{0}}\right) +\int_{{\mathcal{X}}}{\frac{p_{\theta
_{0}}\, p_{\theta _{1}}}{p_{\theta _{2}}}}\,\mathrm{d}\lambda -1
\end{eqnarray*}
where 
\begin{equation*}
\int_{{\mathcal{X}}}{\frac{p_{\theta _{0}}\,p_{\theta _{1}}}{\,p_{\theta
_{2}}}}\,\mathrm{d}\lambda \ =\ \exp \sigma _{0}(\theta _{0},\,\theta
_{1},\,\theta _{2})\text{ \ (cf. (\ref{E.21})).}
\end{equation*}
For $\theta _{0}\in \mathring{\Theta}$\ the desired assertion (\ref{E.18})
follows from here and from the formulas 
\begin{equation*}
D_{0}\left( P_{\theta _{i}},\,P_{\theta _{0}}\right) \ =\ b(\theta
_{i})-b(\theta _{0})-\nabla b(\theta _{0})\,(\theta _{i}-\theta _{0})\text{
\ \ for }i=1,2
\end{equation*}
obtained from (\ref{E.14}).

\medskip

\noindent (c) \ 
The desired formula (\ref{E.19}) follows immediately from 
the definition (\ref{new.1.16b}) and from the formulas 
(\ref{1.15}), (\ref{1.155}), (\ref{E.14}) and (\ref{new.E.14b}).\\
(d) The finally stated invariance is immediate. \hfill $
\square $


\bigskip

The Conclusion 1 of Section III about the relation between scaled Bregman
distances and $\phi $-divergences can be completed by the following relation
between both of them and the classical Bregman distances (\ref{B}).

\bigskip

\paragraph{Conclusion 2}

Let $B_{\phi }(x,y)$\ be the classical Bregman distance (\ref{B}) of
$x,y \in \mathbb{R}^{d}$  and  
$\mathbb{P} = \left\{ P_{\theta }:\theta \in \mathbb{R}^{d}\right\} $
the exponential family with cumulant function $\phi$, i.e., with densities
$p_{\theta }(s)=\exp
\{s\cdot \theta -\phi (\theta)\},$ $s\in \mathbb{R}^{d}$. Then for all
$P_{x}, P_{y}, P_{z} \in \mathbb{P}$
\begin{equation*}
B_{\phi }(x,y)=B_{1}(P_{y},P_{x}|P_{z})=D_{1}(P_{y},P_{x}) \ ,
\end{equation*}
i.e., there is a one-to-one relation
between the classical Bregman distance $B_{\phi }(x,y)$\ and
the scaled Bregman distances $B_{1}(P_{y},P_{x}|P_{z})$ and power
divergences $D_{1}(P_{y},P_{x})$ of the exponential probability measures 
generated by the cumulant function $\phi$.
This means that the family $\left\{ B_{\alpha }(P_{y},P_{x}|P_{z}):\alpha \in \mathbb{R},\text{
}z\in \mathbb{R}^{d}\right\} $\ of scaled Bregman power distances and 
the family $\left\{ D_{\alpha }(P_{y},P_{x}):\alpha
\in \mathbb{R}\right\} $\ of  power divergences extend the classical Bregman distances 
$B_{\phi }(x,y)$ to
which they reduce at $\alpha =1$\ and arbitrary $P_{z} \in \mathbb{P}$. 
In fact, we meet here the extension of the classical Bregman distances
in three different directions:
the first represented by various power parameters $\alpha \in \mathbb{R}$,
the second represented by various possible exponential distributions parametrized by $
\theta \in \mathbb{R}^{d}$, and the third represented by the exponential
distribution parameters $z\in \mathbb{R}^{d}$ which are relevant
when $\alpha \ne 1$.

\bigskip

\paragraph{Remark 4}

We see from Theorems 4 and 5 that -- consistent with (\ref
{new.1.44}), (\ref{1.155}) -- for arbitrary interior parameters $\theta
_{0},\,\theta _{1},\,\theta _{2}\in \mathring{\Theta}$ 
\begin{equation*}
B_{1}\left( P_{\theta _{1}},P_{\theta _{2}}\,|\,P_{\theta _{0}}\right)
=D_{1}\left( P_{\theta _{1}},P_{\theta _{2}}\right) ,
\end{equation*}
i.\thinspace e. that the Bregman distance of order $\alpha =1$ of
exponential family distributions $P_{\theta _{1}},\,P_{\theta _{2}}$ does
not depend on the scaling
distribution  $P_{\theta _{0}}$. The distance of order $
\alpha =0$ satisfies the relation 
\begin{eqnarray*}
& & \hspace{-0.7cm} B_{0}\left( P_{\theta _{1}},P_{\theta _{2}}\,|\,P_{\theta _{0}}\right)
= D_{0}\left( P_{\theta _{1}},P_{\theta _{2}}\right) +\exp \sigma
_{0}(\theta _{0},\,\theta _{1},\,\theta _{2})-1 \notag \\
& &\hspace{-0.7cm} = \ B_{1}\left( P_{\theta _{2}},P_{\theta _{1}}\,|\,P_{\theta _{0}}\right)
+\Delta (\theta _{0},\,\theta _{1},\,\theta _{2})\ ,
\end{eqnarray*}
where 
\begin{equation*}
\Delta (\theta _{0},\,\theta _{1},\,\theta _{2})=\exp \sigma _{0}(\theta
_{0},\,\theta _{1},\,\theta _{2})-1\medskip
\end{equation*}
represents a deviation from the skew-symmetry of the Bregman distances $
B_{0}\left( P_{\theta _{1}},P_{\theta _{2}}\,|\,P_{\theta _{0}}\right) $ and 
$B_{1}\left( P_{\theta _{2}},P_{\theta _{1}}\,|\,P_{\theta _{0}}\right) $\
of $P_{\theta _{1}}$ and $P_{\theta _{2}}$. This deviation is zero if (for
strictly convex $b(\theta )$ if and only if ) $\theta _{0}=\theta _{2}$.

\bigskip

\paragraph{Remark 5}

We see from the formulas 
(\ref{E.8}) -- (\ref{E.19})
that for all $\alpha \in 
\mathbb{R}$ the quantities $D_{\alpha }\left( P_{\theta _{1}},P_{\theta
_{2}}\right) $, $\rho _{\alpha }(\theta _{1},\,\theta _{2})$, $\sigma
_{\alpha }(\theta _{0},\,\theta _{1},\,\theta _{2})$ and $B_{\alpha }\left(
P_{\theta _{1}},P_{\theta _{2}}\,|\,P_{\theta _{0}}\right) $ only depend on
the cumulant function 
$b(\theta)$ defined in (\ref{E.2}), and \textit{not} directly
on the reference measure $\lambda $ used in the definition formulas 
(\ref{E.2}), (\ref{E.1}).


\section{EXPONENTIAL\ APPLICATIONS}

In this section we illustrate the evaluation of scaled Bregman divergences $
B_{\alpha }\left( P_{\theta _{1}},P_{\theta _{2}}\,|\,P_{\theta _{0}}\right) 
$\ for some important discrete and continuous exponential families, and also for
exponentially distributed random processes.

\paragraph{Binomial model} 

Consider for fixed $n\geq 2$ on the observation space $\mathcal{X}
=\{0,...,n\} $ the binomial 
distribution $P_{\theta}$ determined by
\begin{equation}
P_{\theta}[\{x\}] = \lambda[\{x\}] \cdot \exp \{x\cdot \theta
-b(\theta )\} = \binom{n}{x}p^{x}(1-p)^{n-x}  
\notag 
\end{equation}
for $x \in \{0,...,n\}$, where
\begin{equation*}
\lambda[\{x\}] =\binom{n}{x},\text{\ }\theta =\ln \frac{p}{1-p}\in \Theta =
\mathbb{R}\text{\ \ and \ }b(\theta )=n\ln (1+e^{\theta }) \ . 
\end{equation*}
After some calculations one obtains from (\ref{E.10}) and (\ref{sigma})
\begin{equation*}
\rho _{\alpha }(\theta _{1},\theta _{2})=n\ln \frac{1+e^{\alpha \theta
_{1}+(1-\alpha )\theta _{2}}}{(1+e^{\theta _{1}})^{\alpha }(1+e^{\theta
_{2}})^{1-\alpha }}
\end{equation*}
and 
\begin{equation*}
\sigma _{\alpha }(\theta _{0},\theta _{1},\theta _{2})=n\ln \frac{\left(
1+e^{\theta _{1}+(1-\alpha )(\theta _{0}+\theta _{1}-\theta
_{2})}\right) (1+e^{\theta _{2}})^{1-\alpha }}{(1+e^{\theta _{0}})^{\alpha
}(1+e^{\theta _{1}})}.
\end{equation*}
Applying Theorem 5 one achieves an explicit formula for the binomial Bregman distances $B_{\alpha
}\left( P_{\theta _{1}},P_{\theta _{2}}\,|\,P_{\theta _{0}}\right) $\ from
here.

\bigskip

\paragraph{Rayleigh model}

An important role in communication theory play the Rayleigh distributions
defined by the probability densities
\begin{equation}
p_{\theta }(x)=\theta x\exp \left\{ -\frac{\theta x^{2}}{2}\right\} ,\text{
\ }\theta \in \Theta =(0,\infty )  \label{ray}
\end{equation}
with respect to the restriction $\lambda _{+}$ of the Lebesgue measure $
\lambda $\ on the observation space $\mathcal{X}=(0,\infty ).$\ The mapping
\begin{equation*}
T(x)=-\sqrt{2x}
\end{equation*}
from the positive halfline $(0,\infty )$ to the negative halfline $(-\infty
,0)$\ transforms (\ref{ray}) into the family of Rayleigh densities
\begin{eqnarray*}
& & p_{\theta }(x)=\theta \exp \left\{ \theta x\right\} =\exp \left\{ \theta
x-b(\theta )\right\} \notag \\
&& \hspace{4.0cm} \text{ \ for }b(\theta )=-\ln \theta ,\text{ }\theta >0
\end{eqnarray*}
with respect to the restriction $\lambda _{-}$ of the Lebesgue measure $
\lambda $\ on the observation space $\mathcal{X}=(-\infty ,0).$\ These are
the Rayleigh densities in the natural form assumed in (\ref{E.1}). After some
calculations one derives from (\ref{E.10}) 
\begin{equation}
\rho _{\alpha }(\theta _{1},\theta _{2})=\ln \frac{\theta _{1}^{\alpha
}\, \theta _{2}^{1-\alpha }}{\alpha \theta _{1}+(1-\alpha )\theta _{2}}
\label{ro}
\end{equation}
and 
\begin{equation}
\sigma _{\alpha }(\theta _{0},\theta _{1},\theta _{2})=\ln \frac{\theta _{1}
\text{ }\theta _{0}^{1-\alpha }}{\left( \alpha \theta _{1}+(1-\alpha
)(\theta _{0}+\theta _{1}-\theta _{2})\right) \text{ }\theta _{2}^{1-\alpha }
}.  
\notag
\end{equation}
Applying Theorem 5 one obtains the Rayleigh-Bregman distances $B_{\alpha
}\left( P_{\theta _{1}},P_{\theta _{2}}\,|\,P_{\theta _{0}}\right) $\ from
here. \medskip

Theorem 1 about the preservation of the scaled Bregman distances by
statistically sufficient transformations is useful for the evaluation of
these distances in exponential families. It implies for example that these
distances in the normal and lognormal families coincide. The next two
examples dealing with distances of stochastic processes make use of this
theorem too.

\paragraph{Exponentialy distributed signals}

Most of the random processes modelling physical, social and economic
phenomena are exponentially distributed. Important among them are the real
valued L\'{e}vy processes $\boldsymbol{X}\hspace{-0.335cm} \boldsymbol{X}_{t}=(X_{s}:0\leq s\leq t)$ with
trajectories $\boldsymbol{x}\hspace{-0.21cm} \boldsymbol{x}_{t}=(x_{s}:0\leq s\leq t)$ 
from the Skorokchod
observation spaces $(\mathcal{X}_{t},\mathcal{A}_{t})$ and parameters from
the set
\begin{equation*}
\Theta =\{\theta \in \mathbb{R}:c(\theta )<\infty \}
\end{equation*}
defined by means of the function 
\begin{equation*}
c(\theta )=\int_{\mathbb{R}\backslash \{0\}}x^{2}e^{\theta x}/(1+x^{2})
\, \mathrm{d}\nu (x)
\end{equation*}
where $\nu $\ is a L\'{e}vy measure which determines the probability
distribution of the size of jumps of the process and the intensity with
which jumps occur. It is assumed that $0$\ belongs to $\Theta $\ and it is
known (cf., e.g., K\"{u}chler and Sorensen (1994)) that the probability
distributions $P_{t,\theta }$\ induced by these processes on $(\mathcal{X}
_{t},\mathcal{A}_{t})$\ are mutually measure-theoretically equivalent with
the relative densities 
\begin{equation}
\frac{\mathrm{d}P_{t,\theta }}{\mathrm{d}P_{t,0}}(\boldsymbol{x}\hspace{-0.21cm} \boldsymbol{x}_{t})=\exp
\{\theta \text{ }x_{t}-b_{t}(\theta )\}  \label{f}
\end{equation}
for the end $x_{t}$ of the trajectory $\boldsymbol{x}\hspace{-0.21cm} \boldsymbol{x}_{t}$. 
The cumulant function appearing here is
\begin{equation}
b_{t}(\theta )=t\left( \delta \theta +\frac{1}{2}\sigma ^{2}\theta
^{2}+\gamma(\theta )\right)  \label{fff}
\end{equation}
for two genuine parameters $\delta \in \mathbb{R}$ respectively $\sigma >0$ 
of the process
which determine its intensity of drift respectively 
its volatility, and for the function
\begin{equation*}
\gamma(\theta )=\int_{\mathbb{R}\backslash \{0\}}[e^{\theta x}-1-\theta
x/(1+x^{2})]\, \mathrm{d}\nu (x).
\end{equation*}
The formula (\ref{f}) implies that the family $\mathbb{P}_{t}$ $
=\{P_{t,\theta }:\theta \in \Theta \}$\ is exponential on $(\mathcal{X}_{t},
\mathcal{A}_{t})$\ for which the 
  ``extremally reduced'' observation 
$T(\boldsymbol{x}\hspace{-0.21cm} \boldsymbol{x}_{t})=x_{t}$ is statistically sufficient. Thus, by Theorem 1,
\begin{equation}
B(P_{t,\theta _{1}},P_{t,\theta _{2}}|P_{t,0})=B(Q_{t,\theta
_{1}},Q_{t,\theta _{2}}|Q_{t,0})  \label{ff}
\end{equation}
where $Q_{t,\theta }$\ is a probability distribution on the real line governing 
the marginal distribution of the last observed
value $X_{t}$\ of the process $\boldsymbol{X}\hspace{-0.335cm} \boldsymbol{X}_{t}$.

\bigskip 

\paragraph{Queueing processes and Brownian motions}

For illustration of the general result of the  previous subsection we can take the family of 
\textit{Poisson processes} with initial value $X_{0}=0$\ and intensities $
\eta =e^{\theta },\ \theta \in \Theta =\mathbb{R}\ $\ for which $\delta
=\sigma =0$ and $c(\theta )=e^{\theta }-1$ so that $b_{t}(\theta )=t\left(
e^{\theta }-1\right) .$ Then $Q_{t,\theta }$ is the Poisson distribution 
$\text{Poi}(\tau )$ with parameter $\tau =t\eta =te^{\theta }$ and proba\-bilities
\begin{eqnarray*}
& & Q_{t,\theta }[\{x\}] = \frac{e^{-\tau }\left( \tau \right) ^{x}}{x!}=\lambda[\{x\}] \cdot \exp 
\hspace{-0.07cm}\left\{
x\vartheta -e^{\vartheta }\right\} \notag \\
& & \hspace{2.0cm} \text{ \ for \ }\vartheta =\ln \tau
=\theta +\ln t, \ \ \lambda[\{x\}] = \frac{1}{x!} \ .
\end{eqnarray*}
The exponential structure is similar as above, so that 
by applying (\ref{E.10}) to the cumulant function $b(\vartheta
)=e^{\vartheta }=te^{\theta }$\ we get for the Poisson processes with
parameters $\theta _{1}$ and $\theta _{2}$ 
\begin{equation*}
\rho _{\alpha }(\theta _{1},\theta _{2})=t\left[ e^{\alpha \theta
_{1}+(1-\alpha )\theta _{2}}-\alpha e^{\theta _{1}}-(1-\alpha )e^{\theta
_{2}}\right] .
\end{equation*}
Combining this with (\ref{sigma}) and Theorem 5 we obtain 
an explicit formula
for the scaled Bregman distance (\ref{ff}) of these Poisson processes. 

\medskip
To give another illustration of the result of the previous subsection,
let us first introduce the standard Wiener process
$\widetilde{X}_{t}$ which is the L\'{e}vy process
with $\nu \equiv 0$, $\delta=0$, $\sigma=1$ and $\theta=1$.
It defines the \textit{family of Wiener processes} 
\begin{equation}
X_{s}=\theta \, \widetilde{X}_{s},\text{ \ \ }0\leq s\leq t, \ \ \theta \in (0,\infty),  
\notag
\end{equation}
which are L\'{e}vy processes with 
$\delta = 0$, $\sigma =1$ and $c(\theta )\equiv 0$ so that (\ref{fff}) implies $
b_{t}(\theta )=\theta ^{2}/2.$ 
They are well-known models of the random fluctuations 
called Brownian motions. If the initial value $X_{0}$\ is zero then $
Q_{t,\theta }$ is the normal distribution with mean zero and variance $
v^{2}=t\theta ^{2}$. The corresponding Lebesgue densities 
\begin{equation*}
\frac{1}{\sqrt{2\pi v^{2}}}\exp \left\{ -\frac{x^{2}}{2v^{2}}\right\} =\sqrt{
\frac{\vartheta }{\pi }}\exp \left\{ -\vartheta x^{2}\right\} \text{ \ for }
\vartheta =\frac{1}{2v^{2}}\text{\ }
\end{equation*}
are transformed by the mapping $x\longmapsto -\sqrt{|x|}$\ of $\mathbb{R}$\
on the negative halfline $(-\infty ,0)$ into the natural exponential
densities $\exp \left\{ \vartheta x-b(\vartheta )\right\} $ with respect to
the dominating density $1/\sqrt{\pi |x|}$ where $b(\vartheta )=-\frac{1}{2}
\ln \vartheta =-\ln \frac{1}{\theta }+\frac{1}{2}\ln 2t.$ 
Thus by (\ref{E.10})
\begin{equation*}
\rho _{\alpha }(\theta _{1},\theta _{2})= - \ln \frac{\theta _{1}^{\alpha
}\, \theta _{2}^{1-\alpha }}{\alpha \theta _{1}+(1-\alpha )\theta _{2}}
 \qquad (cf.\ (\ref{ro})).
\end{equation*}
This together with (\ref{sigma}) and Theorem 5 leads to the explicit formula
for the scaled Bregman distance (\ref{ff}) of the Wiener processes under
consideration.

\bigskip

\paragraph{Geometric Brownian motions}
From the abovementioned standard Wiener process one can also build up
the \textit{family of geometric Brownian motions} (geometric Wiener processes)
\begin{equation}
Y_{s}=\exp \{\sigma \widetilde{X}_{s}+\theta s\},\text{ \ \ }0\leq s\leq t,
\ \ \theta \in \mathbb{R},
\notag  
\end{equation}
where the family-generating $\theta$ can be interpreted as drift parameters, and 
the volatility parameter $\sigma >0$ 
is assumed to be constant all over the family.
Then, $\sigma \widetilde{X}_{t}+\theta t$\ is normally distributed with mean $m=\theta t$ and
variance $v^{2}=\sigma ^{2}t$,\ and $Y_{t}$\ is lognormally distributed with
the same parameters $m$ and $v^{2}$. By (\ref{ff}), the scaled Bregman
distance of two geometric Brownian motions with parameters $\theta _{1}$,\ $
\theta _{2}$\ reduces to the scaled Bregman distance of two lognormal
distributions LN($\theta _{1}t,\sigma ^{2}t),$ LN($\theta _{2}t,\sigma
^{2}t).$ As said above, it coincides with the scaled Bregman distance of two
normal distributions N($\theta _{1}t,\sigma ^{2}t),$ N($\theta _{2}t,\sigma
^{2}t).$ This is seen also from the fact that the reparametrization
\begin{equation*}
\vartheta =\frac{\mu }{v^{2}},\text{ \ }\tau =\frac{1}{2v^{2}}
\end{equation*}
and transformations $\mathbb{R}\longmapsto \mathbb{R}^{2}$\ similar to that
from the previous example lead in both distributions N($\mu , v^{2}$)
and LN($\mu , v^{2}$) to the same natural exponential density 
\begin{equation*}
p_{\vartheta ,\tau }(x_{1},x_{2})=\exp \left\{ x_{1}\vartheta +x_{2}\tau -
b(\vartheta ,\tau )\right\}
\end{equation*}
with 
\begin{equation*}
b(\vartheta ,\tau )= \frac{1}{2}\, \ln \tau +\frac{\vartheta ^{2}}{4\tau }.
\end{equation*}
These two distributions differ just in the dominating measures on the
transformed observation space $\mathcal{X=}$ $\mathbb{R}^{2}$. 
For
$(\mu_{1},v_{1}^2)=(\theta _{1}t,\sigma ^{2}t)$ \ and \ 
$(\mu_{2},v_{2}^2)=(\theta _{2}t,\sigma ^{2}t)$ 
we get
\begin{equation*}
(\vartheta_1 , \tau_1 ) =  \left(\frac{\theta _{1}}{\sigma ^{2}}\, , \, \frac{1}{2\sigma ^{2}t}\right)
\text{ \ and \ }
(\vartheta_2 , \tau_2 ) =  \left(\frac{\theta _{2}}{\sigma ^{2}}\, , \, \frac{1}{2\sigma ^{2}t}\right)
\end{equation*}
and thus  
\begin{eqnarray*}
&&\hspace{-0.85cm} b(\alpha (\vartheta _{1},\tau _{1})+(1-\alpha )(\vartheta _{2},\tau
_{2}))-\alpha b(\vartheta _{1},\tau _{1})-(1-\alpha )b(\vartheta _{2},\tau
_{2}) \\[0.15cm]
&&\hspace{-0.7cm}= \frac{\left( \alpha \theta _{1}+(1-\alpha )\theta _{2}\right) ^{2}-\alpha
\theta _{1}^{2}+(1-\alpha )\theta _{2}^{2}}{2\sigma ^{2}} \ t \ .
\end{eqnarray*}
Hence, for distributions $P_{t,\theta _{1}},$ $P_{t,\theta _{2}}$\ of the
geometric Brownian motions considered above we get from (\ref{E.10})
\begin{equation*}
\rho _{\alpha }(\theta _{1},\theta _{2})=\frac{\left[ \left( \alpha \theta
_{1}+(1-\alpha )\theta _{2}\right) ^{2}-\alpha \theta _{1}^{2}+(1-\alpha
)\theta _{2}^{2}\right]}{2\sigma ^{2}} \ t \ .
\end{equation*}
The expression (\ref{sigma}) can be automatically evaluated using this.
Applying both these results in Theorem 5 one obtains explicit formula for
the scaled Bregman distance (\ref{ff}) of these geometric Brownian motions.

\section*{Acknowledgment}
We are grateful to all three referees for useful
suggestions.

\ifCLASSOPTIONcaptionsoff
  \newpage
\fi



%


\section*{References}

\begin{description}[\IEEEsetlabelwidth{}]

\small

\item Amari S.-I.\ (2007), \textquotedblleft Integration of stochastic models by minimizing 
$\alpha $-divergence,\textquotedblright\ \textsl{Neural Computation}, vol.\ 19, no.\ 10, 
pp.\ 2780-2796.

\vspace{0.15cm}

\item Banerjee, A., Guo, X., and Wang, H.\ (2005a), \textquotedblleft
On the optimality of conditional expectation as a Bregman predictor,
\textquotedblright\ \textsl{IEEE Transaction on Information theory}, 
vol.\ 51, no.\ 7, pp.\ 2664-2669.

\vspace{0.15cm}

\item Banerjee, A., Merugu, S., Dhillon, I.S.\ and Ghosh, J.\ (2005b), \textquotedblleft
Clustering with Bregman divergences,\textquotedblright\ \textsl{J.\ Machine Learning Research}, 
vol.\ 6, pp.\ 1705-1749.

\vspace{0.15cm}

\item Bartlett, P.L., Jordan M.I. and McAuliffe, J.D.\ (2006), \textquotedblleft Convexity,
classification and risk bounds,\textquotedblright\ \textsl{JASA}, vol.\ 101,pp.\ 138-156.

\vspace{0.15cm}

\item Bauschke, H.H.\ and Borwein, J.M.\ (1997), 
\textquotedblleft Legendre functions and the method of random Bregman projections,
\textquotedblright\ \textsl{J.\ Convex Analysis}, vol.\ 4, No.\ 1, pp.\ 27-67.

\vspace{0.15cm}

\item Boratynska, A.\ (1997), \textquotedblleft Stability of Bayesian inference in exponential
families,\textquotedblright\ \textsl{Statist.\ \& Probab. Letters}, vol.\ 36, pp.\ 173-178.

\vspace{0.15cm}

\item Bregman, L.M.\ (1967), \textquotedblleft The relaxation method of finding the common
point of convex sets and its application to the solution of problems in
convex programming,\textquotedblright\ \textsl{USSR Computational Mathematics and Mathematical
Physics}, vol.\ 7, no.\ 3, pp.\ 200-217.

\vspace{0.15cm}

\item Brown, L.D.\ (1986), \textsl{Fundamentals of Statistical Exponential
Families}. Hayward, California: Inst.\ of Math.\ Statistics.

\vspace{0.15cm}

\item Butnariu, D.\ and Resmerita, E.\ (2006), 
\textquotedblleft Bregman distances, totally convex functions,
and a method for solving operator equations in Banach spaces,
\textquotedblright\ \textsl{Abstr.\ Appl.\ Anal.}, vol.\ 2006, Art.\ ID 84919, 39 pp.

\vspace{0.15cm}

\item Byrne, C.\ (1999), \textquotedblleft Iterative projection onto convex sets using
multiple Bregman distances,\textquotedblright\ \textsl{Inverse Problems}, vol.\ 15, pp.\
1295-1313.

\vspace{0.15cm}

\item Carlson, B.A.\ and Clements, M.A.\ (1991), \textquotedblleft A computationally compact
divergence measure for speech processing.\textit{\ }\textsl{IEEE
Transactions on PAMI}, vol.\ 13, pp.\ 1255-1260.

\vspace{0.15cm}

\item Censor, Y.\ and Zenios, S.A.\ (1997),  \textsl{Parallel Optimization - Theory,
Algorithms, and Applications}. 
New York: Oxford University Press.

\vspace{0.15cm}

\item Cesa-Bianchi, N.\ and Lugosi, G.\ (2006), \textsl{Prediction,
Learning, Games}. Cambridge: Cambridge University Press.

\vspace{0.15cm}

\item Collins, M., Schapire, R.E.\ and Singer, Y.\ (2002), \textquotedblleft Logistic
regression, AdaBoost and Bregman distances,\textquotedblright\ \textsl{Machine Learning}, 
vol.\ 48, pp.\ 253-285.

\vspace{0.15cm}

\item Csisz\'{a}r, I.\ (1963), \textquotedblleft Eine informationstheoretische Un\-gleichung und ihre
Anwendung auf den Beweis der Ergodizit\"at von Markoffschen Ketten.
\textsl{Publ.\ Math.\ Inst.\ Hungar.\ Acad.\ Sci., ser.\ A}, vol.\ 8, pp.\ 85-108.

\vspace{0.15cm}

\item Csisz\'{a}r, I.\ (1967), \textquotedblleft Information-type measures of difference of
probability distributions and indirect observations.
\textsl{Studia Sci.\ Math.\ Hungar.}, vol.\ 2, pp.\ 299-318.

\vspace{0.15cm}

\item Csisz\'{a}r, I.\ (1991), \textquotedblleft Why least squares and maximum entropy? An
axiomatic approach to inference for linear inverse problems,\textquotedblright\ \textsl{Annals
of Statistics}, vol.\ 19, no.\ 4, pp.\ 2032-2066.

\vspace{0.15cm}

\item Csisz\'{a}r, I.\ (1994), \textquotedblleft Maximum entropy and related methods,\textquotedblright\ \textsl{
Trans.\ 12th Prague Conf.\ Information Theory, Statistical Decision
Functions and Random Processes}. Prague, Czech Acad.\ Sci., pp.\ 58-62.

\vspace{0.15cm}

\item Csisz\'{a}r, I.\ (1995), \textquotedblleft Generalized projections for non-negative
functions,\textquotedblright\ \textsl{Acta Mathematica Hungarica}, vol.\ 68, pp.\ 161-186.

\vspace{0.15cm}

\item Csisz\'{a}r, I.\ and Mat\'{u}\v{s}, F.\ (2008), \textquotedblleft On minimization of
entropy functionals under moment constraints,\textquotedblright\ \textsl{Proceedings of ISIT
2008,} Toronto, Canada, pp.\ 2101-2105.

\vspace{0.15cm}

\item Csisz\'{a}r, I.\ and Mat\'{u}\v{s}, F.\ (2009), \textquotedblleft On minimization of
multivariate entropy functionals,\textquotedblright\ \textsl{Proceedings of ITW 2009,} Volos,
Greece, pp.\ 96-100.

\vspace{0.15cm}

\item Do, M.N.\ and Vetterli, M.\ (2002), \textquotedblleft Wavelet-based texture retrieval
using generalized Gaussian density and Kullback-Leibler distance,\textquotedblright\ \textsl{
IEEE\ Transactions on Image Processing}, vol.\ 11, pp.\ 146-158.

\vspace{0.15cm}

\item Freund, Y.\ and Schapire, R.E.\ (1997), \textquotedblleft A decision-theoretic
generalization of on-line learning and an application to boosting,\textquotedblright\ \textsl{
J. Comput. Syst. Sci.}, vol.\ 55, pp.\ 119-139.

\vspace{0.15cm}

\item Halmos, P.R.\ (1964), \textsl{Measure Theory}. New York: D.\ Van Nostrand.

\vspace{0.15cm}

\item Hertz, T., Bar-Hillel, A.\ and Weinshall, D.\ (2004), \textquotedblleft Learning distance
functions for information retrieval,\textquotedblright\ in \textsl{Proc.\ IEEE Comput.\ Soc.\ Conf.\ on Computer Vision
and Pattern Rec.\ CVPR}, vol.\ 2, II-570 - II-577.

\vspace{0.15cm}

\item Jones, L.K.\ and Byrne, C.L.\ (1990), \textquotedblleft General entropy criteria for
inverse problems, with applications to data compression, pattern
classification, and cluster analysis,\textquotedblright\ \textsl{IEEE Trans.\ Inform.\ Theory} 
vol.\ 36, no.\ 1, pp.\ 23-30.

\vspace{0.15cm}

\item K\"{u}chler, U.\ and Sorensen, M.\ (1994), \textquotedblleft Exponential families of
stochastic processes and L\'{e}vy processes,\textquotedblright\ \textsl{J. of Statist. Planning
and Inference}, vol.\ 39, pp.\ 211-237.

\vspace{0.15cm}

\item Lehman, E.L.\ and Romano J.P.\ (2005), \textsl{Testing Statistical
Hypotheses.} Berlin: Springer.

\vspace{0.15cm}

\item Liese, F.\ and Vajda, I.\ (1987),  \textsl{Convex Statistical Distances}. 
Leipzig: Teubner.

\vspace{0.15cm}

\item Liese, F.\ and Vajda, I.\ (2006), \textquotedblleft On divergences and informations in
statistics and information theory,\textquotedblright\ \textsl{IEEE Transaction on Information
theory}, vol.\ 52, no.\ 10, pp.\ 4394-4412.

\vspace{0.15cm}

\item Lloyd, S.P.\ (1982), \textquotedblleft Least squares quantization in PCM,\textquotedblright\ \textsl{IEEE
Transactions on Inform. Theory}, vol.\ 28, no.\ 2, pp.\ 129-137.

\vspace{0.15cm}

\item Marquina, A.\ and Osher, S.J.\ (2008), \textquotedblleft Image super-resolution by
TV-regularization and Bregman iteration,\textquotedblright\ \textsl{J.\ Sci.\ Comput.}, vol.\ 
37, pp.\ 367-382.

\vspace{0.15cm}

\item Murata, N., Takenouchi, T., Kanamori, T. and Eguchi, S.\ (2004), \textquotedblleft
Information geometry of $\mathcal{U}$-Boost and Bregman divergence,\textquotedblright\ \textsl{
Neural Computation}, vol.\ 16, no.\ 7, pp.\ 1437-1481.

\vspace{0.15cm}

\item Nock, R.\ and Nielsen, F.\ (2009), \textquotedblleft Bregman divergences and surrogates
for learning,\textquotedblright\ \textsl{IEEE Transactions on PAMI},
vol.\ 31,  no.\ 11, pp.\ 2048 - 2059.

\vspace{0.15cm}

\item Pardo, M.C.\ and Vajda, I.\ (1997), \textquotedblleft About distances of discrete
distributions satisfying the data processing theorem of information theory,\textquotedblright\ 
\textsl{IEEE Transaction on Information theory}, vol.\ 43, no.\ 4, pp.\
1288-1293.

\vspace{0.15cm}

\item Pardo, M.C.\ and Vajda, I.\ (2003), \textquotedblleft On asymptotic properties of
information-theoretic divergences,\textquotedblright\ \textsl{IEEE Transaction on Information
theory}, vol.\ 49, no.\ 7, pp.\ 1860-1868.

\vspace{0.15cm}

\item R\'{e}nyi, A.\ (1961), \textquotedblleft On measures of entropy and information,\textquotedblright\ in \textsl{
Proc.\ 4th Berkeley Symp.\ Math.\ Stat.\ Probab.}, vol.\ 1, pp.\ 547-561. Berkeley, CA: Univ. of California Press.

\vspace{0.15cm}

\item Resmerita, E.\ and Anderssen R.S.\ (2007), \textquotedblleft Joint additive
Kullback-Leibler residual minimization and regularization for linear inverse
problems,\textquotedblright\ \textsl{Math.\ Meth.\ Appl.\ Sci.}, vol.\ 30, no.\ 13, pp.\ 1527-1544.

\vspace{0.15cm}

\item Scherzer, O., Grasmair, M., Grossauer, H., Haltmeier, M.\ and Lenzen, F.\ (2008);
\textsl{Variational methods in imaging}. New York: Springer.

\vspace{0.15cm}

\item Stummer, W.\ (2007), \textquotedblleft Some Bregman distances between financial
diffusion processes,\textquotedblright\ \textsl{Proc.\ Appl.\ Math.\ Mech.}, vol.\ 7, no.\ 1,
pp.\ 1050503 - 1050504. 

\vspace{0.15cm}

\item Stummer, W.\ and Vajda, I.\ (2010), \textquotedblleft On divergences of finite measures
and their applicability in statistics and information theory,\textquotedblright\ \textsl{
Statistics}, vol.\ 44, pp.\ 169-187.

\vspace{0.15cm}

\item Teboulle, M.\ (2007), \textquotedblleft A unified continuous optimization framework for
center-based clustering methiods,\textquotedblright\ \textsl{Journal of Machine Learning
Research}, vol.\ 8, pp.\ 65-102.

\vspace{0.15cm}

\item Vajda, I.\ (2009), \textquotedblleft \ On metric divergences of probability measures,\textquotedblright\ 
\textsl{Kybernetika}, vol.\ 45, no.\ 5 (in print).

\vspace{0.15cm}

\item Vajda, I.\ and Zv\'{a}rov\'{a}, J. (2007), \textquotedblleft On generalized entropies,
Bayesian decisions and statistical diversity,\textquotedblright\ 
\textsl{Kybernetika}, vol.\ 43, no.\ 5, pp. 675-696.

\vspace{0.15cm}

\item Veldhuis, R.N.J.\ (2002), \textquotedblleft The centroid of the Kullback-Leibler
distance,\textquotedblright\ \textsl{IEEE Signal Processing Letters}, vol.\ 9, no.\ 3, pp.\ 96-99.

\vspace{0.15cm}

\item Xu, J.\ and Osher, S.\ (2007), \textquotedblleft Iterative regularization and nonlinear
inverse scale space applied to wavelet-based denoising,\textquotedblright\ \textsl{IEEE
Transaction on Image Processing}, vol.\ 16, no.\ 2, pp.\ 534-544.

\end{description}


\vspace{0.2cm}

\textbf{Wolfgang Stummer} graduated from the Johannes Kepler University Linz,
Austria, in 1987 and received the Ph.D.\ degree in 1991 from the University of Zurich,
Switzerland.\\
\indent
From 1993 to 1995 he worked as Research Assistant at the University of London and the
University of Bath (UK). From 1995 to 2001 he was Assistant Professor
at the University of Ulm (Germany). From 2001 to 2003 he held a 
Term Position as a Full Professor at the University of Karlsruhe (now KIT; Germany)
where he continued as Associate Professor until 2005.
Since then, he is affiliated as Full Professor at the Department of Mathematics, 
University of Erlangen-N\"urnberg FAU (Germany); at the latter, he is also a Member of the
School of Business and Economics.

\vspace{0.2cm}

\textbf{Igor Vajda} (M'90 -– F'01) was born in 1942 and passed away suddenly after a
short illness on May 2, 2010. He graduated from the Czech Technical University,
Czech Republic, in 1965 and received the Ph.D. degree in 1968 from Charles
University, Prague, Czech Republic.\\
\indent He worked at UTIA (Institute of Information Theory and Automation,
Czech Academy of Sciences) from his graduation until his death, and became
a member of the Board of UTIA in 1990. He was a visiting professor at
the Katholieke Universiteit Leuven, Belgium; the Universidad Complutense
Madrid, Spain; the Universit\'e de Montpellier, France; and the Universidad
Miguel H\'ernandez, Alicante, Spain. He published four monographs and more
than 100 journal publications.\\
\indent Dr.\, Vajda received the Prize of the Academy of Sciences, the Jacob Wolfowitz
Prize, the Medal of merits of Czech Technical University, several Annual prizes
from UTIA, and, posthumously, the Bolzano Medal from the Czech Academy
of Sciences.

\end{document}